\documentclass[runningheads]{llncs}

\usepackage[year=2024]{eccv}

\usepackage{eccvabbrv}

\usepackage{graphicx,float}
\usepackage[dvipsnames]{xcolor}
\usepackage{booktabs}
\usepackage[mathscr]{euscript}
\usepackage{amsmath}
\usepackage{amssymb}
\usepackage{algpseudocode}
\usepackage{algorithm,colortbl}
\definecolor{Gray}{gray}{0.82}
\definecolor{LightGray}{gray}{0.88}
\usepackage[accsupp]{axessibility}  
\usepackage{wrapfig}

\usepackage[pagebackref,breaklinks,colorlinks,citecolor=eccvblue]{hyperref}

\usepackage{orcidlink}

\begin{document}

\title{Group Testing for Accurate and Efficient Range-Based Near Neighbor Search for Plagiarism Detection} 

\titlerunning{Group Testing for NN Search}

\author{Harsh Shah\inst{1}\orcidlink{0009-0008-0688-9433} \and
Kashish Mittal\inst{1}\orcidlink{0009-0004-7003-6444} \and
Ajit Rajwade\inst{1}\orcidlink{0000-0001-6463-3315}}

\authorrunning{H.~Shah et al.}

\institute{Department of Computer Science \& Engineering, Indian Institute of Technology Bombay}
\maketitle

\begin{abstract}
This work presents an adaptive group testing framework for the range-based high dimensional near neighbor search problem. Our method efficiently marks each item in a database as neighbor or non-neighbor of a query point, based on a cosine distance threshold without exhaustive search. Like other methods for large scale retrieval, our approach exploits the assumption that most of the items in the database are unrelated to the query. Unlike other methods, it does not assume a large difference between the cosine similarity of the query vector with the least related neighbor and that with the least unrelated non-neighbor. Following a multi-stage adaptive group testing algorithm based on binary splitting, we divide the set of items to be searched into half at each step, and perform dot product tests on smaller and smaller subsets, many of which we are able to prune away. We experimentally show that, using softmax-based features, our method achieves a more than ten-fold speed-up over exhaustive search with no loss of accuracy, on a variety of large datasets. 
Based on empirically verified models for the distribution of cosine distances, we present a theoretical analysis of the expected number of distance computations per query and the probability that a pool with a certain number of members will be pruned. 
Our method has the following features: (\textit{i}) It implicitly exploits useful distributional properties of cosine distances unlike other methods; (\textit{ii}) All required data structures are created purely offline; (\textit{iii}) It does not impose any strong assumptions on the number of true near neighbors; (\textit{iv}) It is adaptable to streaming settings where new vectors are dynamically added to the database; and (\textit{v}) It does not require any parameter tuning. The high recall of our technique makes it particularly suited to plagiarism detection scenarios where it is important to report every database item that is sufficiently similar item to the query.
\end{abstract}

\section{Introduction}
Near neighbor (NN) search is a fundamental problem in machine learning, computer vision, image analysis and recommender systems. Consider a database $\mathcal{D}$ of $N$ vectors given as $\{\boldsymbol{f_1}, \boldsymbol{f_2}, \cdots, \boldsymbol{f_N}\}$ where each $\boldsymbol{f_i} \in \mathbb{R}^d$. Given a query vector $\boldsymbol{q} \in \mathbb{R}^d$, our aim is to determine all those 
vectors in $\mathcal{D}$ such that $\textrm{sim}(\boldsymbol{q},\boldsymbol{f_i}) \geq \rho$, where $\textrm{sim}(.,.)$ denotes some similarity measure, and $\rho$ denotes some threshold on $\textrm{sim}$. This is called a similarity measure based NN query. Another variant is the $K$ nearest neighbor (KNN) search where the aim is to find the $K$ most similar neighbors to $\boldsymbol{q}$. 

A significant portion of the literature on NN search focusses on \emph{approximate} methods, often resulting in less than perfect recall rates, i.e. many database images that are similar to the query image are not retrieved. This deficiency renders them unsuitable for applications requiring high recall, such as plagiarism detection. In such scenarios, the query vectors, such as images or music files, are compared to an existing database of similar entities, and \emph{all} vectors exhibiting high similarity with the query need to be retrieved for plagiarism assessment. The failure to retrieve a near neighbor could potentially lead to unfair evaluation of submissions to photography or music competitions, or online forums.

\noindent\textbf{Related Work on NN Search:} There exists a large body of literature on approximate NN search in higher dimensions. One of the major contributions in this area is called Locality Sensitive Hashing (LSH) \cite{Indyk1998}.
Here, a hashing function partitions the dataset into buckets, such that only close vectors are likely to be hashed to the same bucket. The LSH family of algorithms consists of many variants such as those involving multiple buckets per query vector, data-driven or machine learning based construction of LSH functions \cite{Andoni2015,Dong2019}, and count-based methods \cite{Lv2007}. In general, LSH based methods are difficult to implement for very high dimensions owing to the curse of dimensionality in constructing effective hash tables. 
Dimensionality reduction algorithms can mitigate this issue to some extent, but they are themselves expensive to implement and incur some inevitable data loss. There is also significant research in graph-based methods for nearest neighbor (NN) search, as explored in \cite{Malkov2018, Fu2019, graphknn}. These methods typically achieve high accuracy and efficient query times. However, they often need relatively high pre-processing times, making them less suitable for dynamic datasets that require frequent addition of new points. 
In addition, various randomized variants of KD-trees and priority search K-means trees have been investigated for efficient approximate NN search in high-dimensional spaces, as seen in \cite{Ram2019} and \cite{Muja2014}. Despite their efficiency, these approaches involve numerous parameters, which can complicate tuning, especially for large datasets.

\noindent\textbf{Overview of Group Testing:} In this work, we present a group testing (GT) based approach for the NN problem. We start by presenting a brief overview of the field of GT. Consider $N$ items $\{x_i\}_{i=1}^N$ (also called `samples' in many references), out of which only a small number $s \ll N$ are `defective'. We consider the $i$th item to be defective if $x_i = 1$ and non-defective if $x_i = 0$. GT involves testing a certain number of `\textbf{pools}' ($m \ll N$) to identify the defective items from $\{x_i\}_{i=1}^N$. Each pool (also called `group') is created by mixing together or combining subsets of the $N$ items. Thus the $j$th pool ($j \in [m]$) is represented as $y_j = \bigvee_{i=1}^N A_{ji} x_i$,   
where $\bigvee$ stands for bit-wise OR, and $\boldsymbol{A} \in \{0,1\}^{m \times N}$ is a binary pooling matrix where $A_{ji} = 1$ if the $i$th item participates in the $j$th pool and 0 otherwise. GT has a rich literature dating back to the seminal `Dorfman's method' \cite{Dorfman1943} which is a popular two-round testing strategy. In the first round of Dorfman's method, some $N/K$ pools, each consisting of $K$ items, are tested. The items that participated in pools that test \emph{negative} (i.e., pools that are deemed to contain no defective item) are declared non-defective, whereas all items that participated in the \emph{positive} pools (i.e., pools that are deemed to contain at least one defective item) are tested individually in the second round. Theoretical analysis shows that for $s \ll N$, this method requires much fewer than $N$ tests in the worst case. This algorithm was extended to some $r$ rounds in \cite{Li1962}, where positive groups are further divided into smaller groups, and individual testing is performed only in the $r$th round. The methods in \cite{Dorfman1943, Li1962} are said to be \textbf{adaptive} in nature, as the results of one round are given as input to the next one. The GT literature also contains a large number of \textbf{non-adaptive} algorithms that make predictions in a single round.  A popular algorithm is called \textsc{Comp} or combinatorial orthogonal matching pursuit, which declares all items participating in a negative pool to be negative, and all the rest to be positive \cite{Aldridge2019}. Many variants of \textsc{Comp} such as Noisy \textsc{Comp} (\textsc{NComp}) or Definite Defectives (\textsc{Dd}) are also popular \cite{Aldridge2019}. A popular class of non-adaptive GT algorithms \cite{ghosh2021compressed} draw heavily from the compressed sensing (CS) literature \cite{Candes2008,Vidyasagar2019}. Such algorithms can consider quantitative information for each $\{x_i\}_{i=1}^N$ so that each $x_i \in \mathbb{R}$, and every pool $y_j$ also gives a real-valued result. The pooling matrix $\boldsymbol{A}$ remains binary, and we have the relation
$\forall j \in [m], y_j  = \sum_{i=1}^N A_{ji} x_i$. Given $\{y_j\}_{j=1}^m$ and $\boldsymbol{A}$, the unknown vector $\boldsymbol{x}$ can be recovered using efficient convex optimization methods such as the \textsc{Lasso} given by $\boldsymbol{\hat{x}} = \textrm{Argmin}_{\boldsymbol{x}} \|\boldsymbol{y}-\boldsymbol{Ax}\|^2_2 + \lambda \|\boldsymbol{x}\|_1$, 
where $\lambda$ is a regularization parameter \cite{Hastie2015}. As proved in \cite{Hastie2015, Davenport2012}, the \textsc{Lasso} estimator produces provably accurate estimates of $\boldsymbol{x}$ given a sufficient number of measurements (or pools) in $\boldsymbol{y}$ and a carefully designed matrix $\boldsymbol{A}$.

\noindent\textbf{Overview of our work:} GT algorithms have been applied to the NN search problem recently in  \cite{Shi2014, Iscen2018, Engels2021}. However in these papers, the nearest neighbor queries are of an \emph{approximate} nature, i.e. the results may contain a number of vectors which are not near neighbors of the query vector $\boldsymbol{q}$, and/or may miss some genuine near neighbors. On the other hand, we focus on \emph{accurate} query results with good computational efficiency in practice. Our approach is based on an adaptive GT approach called binary splitting. Like existing GT based algorithms such as \cite{Shi2014, Iscen2018, Engels2021}, our approach is also based on the assumption that in high-dimensional NN search, most vectors in the database $\mathcal{D}$ are dissimilar to a query vector $\boldsymbol{q}$, and very few vectors qualify as near neighbors. However, our method does not require the distance between the most unrelated neighbor and the most closely related non-neighbor to be large, unlike \cite{Engels2021}. Additionally, many existing approaches \cite{Engels2021,Iscen2018,Shi2014,Muja2014,Andoni2015,pham2022,Guo2020,Faiss,Babenko2014} require parameter tuning in a manner that is not fully data driven. It is also a tedious process for large databases, particularly when dealing with algorithms with large index building times. Moreover, algorithms with large index building times or those unable to efficiently handle the addition of new points are ill-suited for streaming settings. In this work, we present a range-based near neighbor search algorithm in high dimensions built on a group testing framework that has the following appealing properties for softmax-based feature descriptors: (\textit{i}) low index building time, (\textit{ii}) perfect precision and recall without exhaustive search, (\textit{iii}) no parameter tuning, (\textit{iv}) low time complexity ($\mathcal{O}(1)$) for addition of new vectors making the algorithm suitable for streaming tasks, and (\textit{v}) a unique method of incorporating distributional properties of the similarity values between query and database vectors into the search process. 

\noindent\textbf{Organization of the paper:} The core method is presented in Sec.~\ref{sec:method}. We present detailed analytical comparisons with existing GT-based techniques for NN search in Sec.~\ref{sec:comparison}. Experimental results and comparisons to a large number of recent NN search methods are presented in Sec.~\ref{sec:results}. A theoretical analysis of our approach is presented in Sec.~\ref{sec:theoretical_analysis}. We conclude in Sec.~\ref{sec:conclusion}.

\section{Description of Method}
\label{sec:method}
In our technique, a pool of high-dimensional vectors is created by simply adding together the participating vectors, which may for example be (vectorized) images or features extracted from them. Consider a query vector $\boldsymbol{q} \in \mathbb{R}^d$ and the database $\mathcal{D}$ consisting of $\{\boldsymbol{f_1}, \boldsymbol{f_2},\cdots,\boldsymbol{f_N}\}$ where each $\boldsymbol{f_i} \in \mathbb{R}^d$. Throughout this paper, we consider the similarity measure $\textrm{sim}(\boldsymbol{q}, \boldsymbol{f_i}) \triangleq \boldsymbol{q}^t \boldsymbol{f_i}$. For simplicity, we consider the query vector $\boldsymbol{q}$ as well as the vectors in $\mathcal{D}$ to be element-wise non-negative. This constraint is satisfied by images or their features based on ReLU or softmax non-linearities (see the end of this section for a method to handle negative-valued descriptors). We also consider that the vectors in question have unit magnitude, without any loss of generality. Due to this, $\boldsymbol{q}^t \boldsymbol{f_i}$ acts as a cosine similarity measure. 

Our aim is to retrieve all `defective' members of $\mathcal{D}$, i.e. each member $\boldsymbol{f_i}$ for which $\boldsymbol{q}^t \boldsymbol{f_i} \geq \rho$, where $\rho$ is some pre-specified threshold. With this aim in mind, we aim to efficiently prune away any vector $\boldsymbol{f_i}$ from $\mathcal{D}$ for which it is impossible to have $\boldsymbol{q}^t \boldsymbol{f_i} \geq \rho$. Vector pooling using summations proves to be beneficial for this purpose. Consider the toy example of a pool $\boldsymbol{y} = \boldsymbol{f_1} + \boldsymbol{f_2} + \boldsymbol{f_3} + \boldsymbol{f_4}$. If $\boldsymbol{q}^t \boldsymbol{y} < \rho$ (i.e. the pooled result is negative), then one can clearly conclude that $\boldsymbol{q}^t \boldsymbol{f_i} < \rho$ for each $i \in \{1,2,3,4\}$. However if  $\boldsymbol{q}^t \boldsymbol{y} \geq \rho$, i.e. it yields a positive result, then we need to do further work to identify the defective members (if any). For this, we split the original pool and create two pools $\boldsymbol{y_1}$ and $\boldsymbol{y_2}$, each with approximately half the number of members of the original pool. We recursively compute $\boldsymbol{q}^t \boldsymbol{y_2}$, and then obtain $\boldsymbol{q}^t \boldsymbol{y_1} = \boldsymbol{q}^t \boldsymbol{y} - \boldsymbol{q}^t \boldsymbol{y_2}$ (where both $\boldsymbol{q}^t \boldsymbol{y}$ and $\boldsymbol{q}^t \boldsymbol{y_2}$ have already been computed earlier, and hence $\boldsymbol{q}^t \boldsymbol{y_1}$ is directly obtained from the difference between them). Thereafter, we proceed in a similar fashion for both branches. 

We now consider the case of searching within the entire database $\mathcal{D}$ defined earlier. We initiate this recursive process with a pool $\boldsymbol{y_0}$ that is obtained from the summation of all $N$ vectors. If $\boldsymbol{y_0}$ yields 
a positive result, we test the query vector with two pools: $\boldsymbol{y_{11}}$ obtained from a summation of vectors $\boldsymbol{f_1}$ through to $\boldsymbol{f_{\lfloor N/2 \rfloor}}$, and $\boldsymbol{y_{12}}$ obtained from a summation of vectors $\boldsymbol{f_{\lfloor N/2 \rfloor+1}}$ through to $\boldsymbol{f_N}$. If $\boldsymbol{y_{11}}$ tests negative, then all its members are declared negative and no further tests are required on its members, which greatly reduces the total number of tests. If $\boldsymbol{y_{11}}$ tests positive, then it is split into two pools 
$\boldsymbol{y_{21}}, \boldsymbol{y_{22}}$ with roughly equal number of members. Tests are carried out further on these two pools recursively. The same treatment is accorded to $\boldsymbol{y_{12}}$, which would be split into pools $\boldsymbol{y_{23}}, \boldsymbol{y_{24}}$ if it were to test positive. This recursive process stops when the pools obtained after splitting contain just a single vector each, in which case each such single vector is tested against $\boldsymbol{q}$. Clearly, the depth of this recursive process is $\lceil \log_2 N \rceil$ given $N$ vectors in $\mathcal{D}$. This procedure is called \textbf{binary splitting}, and it is a multi-round adaptive group testing approach, applied here to NN search. For efficiency of implementation, it is important to be able to create the different pools such as $\boldsymbol{y_0}, \boldsymbol{y_{11}}, \boldsymbol{y_{12}}, \boldsymbol{y_{21}}, \boldsymbol{y_{22}}, \boldsymbol{y_{23}}, \boldsymbol{y_{24}}$, etc., efficiently. For this purpose, we maintain cumulative sums of the following form in memory:
\begin{eqnarray}
\boldsymbol{\widetilde{f}_1} = \boldsymbol{f_1}, 
\boldsymbol{\widetilde{f}_2} = \boldsymbol{\widetilde{f}_1} + \boldsymbol{f_2}, 
...,\boldsymbol{\widetilde{f}_N} = \boldsymbol{\widetilde{f}_{N-1}} + \boldsymbol{f_N}. 
\label{eq:cumsums}
\end{eqnarray}
The cumulative sums are useful for efficient creation of various pools. They are created purely \emph{offline}, independent of any query vector. The pools are created using simple vector difference operations, given these cumulative sums. For example, notice that $\boldsymbol{y_0} = \boldsymbol{\widetilde{f}_N}, \boldsymbol{y_{11}} = \boldsymbol{\widetilde{f}_{\lfloor N/2 \rfloor}}$ and $\boldsymbol{y_{12}} = \boldsymbol{\widetilde{f}_N} - \boldsymbol{\widetilde{f}_{\lfloor N/2 \rfloor}}$. The overall binary splitting procedure is presented in Alg.~\ref{alg:binsplt} (a schematic diagram is provided in \cite{suppmat}). For the sake of greater efficiency, it is implemented non-recursively using a simple stack data structure. 
\begin{algorithm}[h!]
\caption{Iterative Binary Splitting (see \cite[Fig. 1]{suppmat} for a diagram)}\label{alg:binsplt}
\begin{algorithmic}[1]
\State $Fs \gets [\boldsymbol{\widetilde{f}_1}, \boldsymbol{\widetilde{f}_2},...,\boldsymbol{\widetilde{f}_N}]$
\State $S \gets$ stack for storing\ triples $\{si, ei, sim\}$ where $si, ei$ are start and end index for a pool, $sim =$ dot product between query vector $\boldsymbol{q}$ and pool vector. 
\State $Flg \gets$ array of size $N$ flags, all\ 0's initially ($0 =$ non-neighbor, $1 =$ neighbor of $\boldsymbol{q}$)
\State $S.push\ \{1, N, \boldsymbol{q}^t\boldsymbol{\widetilde{f}_N}\}$, i.e. $si = 1, ei = N$
\While{$S.size() \neq 0$} 
    \State $\{si, ei, sim\} \gets S.top()$; $S.pop()$
    \If{$sim \geq \rho$}
        \State $n \gets ei-si+1$
        \If{$n > 2$}
            \State $rsim \gets \boldsymbol{q}^t(\boldsymbol{\widetilde{f}_{ei}} - \boldsymbol{\widetilde{f}_{si+\lfloor n/2 \rfloor-1}})$ \Comment{testing right-half}
            \State $S.push\ \{si+\lfloor n/2 \rfloor, ei, rsim\}$; \Comment{push right branch onto stack}
            \State $S.push\ \{si, si+\lfloor n/2 \rfloor-1,sim-rsim\}$; \Comment{push left branch onto stack}
        \Else
            \If{$n = 2$}
                \State $rsim \gets \boldsymbol{q}^t(\boldsymbol{\widetilde{f}_{ei}} - \boldsymbol{\widetilde{f}_{si}})$
                \State \textbf{set} $Flg[ei] = 1$, \textbf{if} $rsim \geq \rho$
                \State \textbf{set} $Flg[si] = 1$, \textbf{if} $sim - rsim$ $\geq$ $\rho$
            \Else
                \State $Flg[si] = 1$ \Comment{Mark $si$ a neighbor}
            \EndIf
        \EndIf
    \EndIf
\EndWhile
\end{algorithmic}
\end{algorithm}
This binary splitting approach is quite different from Dorfman's algorithm \cite{Dorfman1943}, which performs individual testing in the second round. Our approach is more similar to Hwang's generalized binary splitting approach, a GT procedure which was proposed in \cite{Hwang1972} (see also \cite{gbs_wiki}, Alg. 1.1 of \cite{Aldridge2019}). However compared to Hwang's technique, our approach allows for creation of all pools either in a purely offline manner or with simple updates involving a single vector-subtraction operation (for example, $\boldsymbol{y_{12}} = \boldsymbol{\widetilde{f}_N} - \boldsymbol{\widetilde{f}_{\lfloor N/2 \rfloor}}$ in the earlier example). On the other hand, the method in \cite{Hwang1972} detects defectives one at a time, after going through all $\log_2 N$ levels of recursion. Once a defective item is detected, it is discarded and all pools are created again. The entire set of $\log_2 N$ levels of recursion are again executed. This process is repeated until all defectives have been detected and discarded one at a time. The method from \cite{Hwang1972} is computationally inefficient for high-dimensional NN search, as the pools must be created afresh for each level of recursion. This is because if a vector at index $i$ is discarded, all cumulative sums from index $i+1$ to $N$ need to be updated. Such updates are not required in our method. We also emphasize that the techniques in \cite{Dorfman1943,Aldridge2019,Hwang1972,Li1962} have all been applied \emph{only} for binary GT and \emph{not} for NN search (see the beginning of Sec.~\ref{sec:theoretical_analysis} for an important difference between binary GT and NN search). 
The binary splitting procedure saves on a large number of tests because in high dimensional search, we observe that most vectors in $\mathcal{D}$ are dissimilar to a given query vector $\boldsymbol{q}$, as will be demonstrated in Sec.~\ref{sec:results}. After the first round where pool $\boldsymbol{y_0}$ is created, all other pools are essentially created randomly (since the initial arrangement of the vectors in $\mathcal{D}$ is random). As a result, across the different rounds, many pools test negative and are dropped out, especially in later rounds. In particular, we note that there is no loss of accuracy in this method, although we save on the number of tests by a factor more than ten on some large datasets (as will be shown in Sec.~\ref{sec:results}). 

\noindent\textbf{Pooling using Element-wise Maximum:}
We have so far described pool creation using vector summation. However, pools can also be created by computing the element-wise maximum of all vectors participating in the pool. Given a subset of vectors $\{\boldsymbol{f_k}\}_{k=1}^K$, such a pool $\boldsymbol{y_m}$ is given by: $\forall j \in \{1,2,...,d\}, y_{m,j} = \textrm{max}_j(f_{k,j})$ where $f_{k,j}$ stands for the $j$th element of $\boldsymbol{f_k}$. We denote this operation as $\boldsymbol{y_m} = \textrm{max}(\{\boldsymbol{f_k}\}_{k=1}^K)$. If $\boldsymbol{q}^t \boldsymbol{y_m} < \rho$, it follows that $ \boldsymbol{q}^t \boldsymbol{f_k} < \rho$ for every $\boldsymbol{f_k}$ that contributed to $\boldsymbol{y_m}$. Hence pools created using an element-wise maximum are also well suited for iterative binary splitting in a similar fashion as in Alg.~\ref{alg:binsplt}. 
Separate element-wise maximum vectors need to be stored in memory for each pool. In case of negative values in the query and/or database vectors, the element-wise maximum can be replaced by an element-wise minimum for every index $j$ for which $q_j < 0$. Unlike the summation technique, this handles negative-valued descriptors, albeit at the cost of more memory in order to store both element-wise maximum and element-wise minimum values for each pool.  

\section{Comparison to other Group Testing Methods for NN Search}
\label{sec:comparison}
GT algorithms of different flavors have been applied to the NN search problem in \cite{Shi2014,Iscen2018,Engels2021}. A detailed description of these techniques is presented in the supplemental material at \cite[Sec. 2]{suppmat}. Compared to these three afore-mentioned techniques, our method is exact, with the same accuracy as exhaustive search. The methods \cite{Shi2014,Iscen2018,Engels2021} require choice of various hyper-parameters ($R$, the number of backpropagation steps, for \cite{Shi2014}; sparse coding parameters in \cite{Iscen2018}; bloom filter and hash function parameters ($B, R, m, L$) in \cite{Engels2021} -- see \cite[Sec. 2]{suppmat}) for which there is no clear data-driven selection procedure. Our method, however, requires no parameter tuning. In experiments, we have observed a speed up of more than ten-fold in querying time with our method as compared to exhaustive search on some datasets. Like \cite{Shi2014}, and unlike \cite{Engels2021, Iscen2018}, our method does have a large memory requirement, as we require all pools to be in memory. Some techniques such as \cite{Engels2021} require the nearest neighbors to be significantly more similar than all other members of $\mathcal{D}$ (let us call this condition $\mathscr{C}1$). Our method does not have such a requirement. However, our method will perform more efficiently for queries for which a large number of similarity values turn out to be small, and only a minority are above the threshold $\rho$ (let us call this condition $\mathscr{C}2$). In our experimental study on diverse datasets, we have observed that $\mathscr{C}2$ is  true always. On the other hand, $\mathscr{C}1$ does not hold true in a large number of cases, as also reported in \cite[Sec. 5.3]{Engels2021}.

\noindent We also experimented with applying the latest developments in the compressed sensing (CS) literature \cite{Vidyasagar2019} (which is closely related to GT) to the \emph{approximate} NN search problem, albeit unsuccessfully. The theoretical reasons for the failure of the latest CS techniques for NN search are described in \cite[Sec. 3]{suppmat}. 

\section{Experimental Results}
\label{sec:results}
In this section, we present experimental results conducted in two settings: (\textit{i}) Non-streaming setting with a static database, and (\textit{ii}) Streaming setting with incremental additions of new vectors to the database. In both the settings, we tested our algorithms on the following publicly available image datasets, of which the first two are widely used in retrieval and NN search problems: (\textit{i}) \textsf{MIRFLICKR}, a set of 1 million (1M) images from the MIRFLICKR dataset (subset of images from Flickr) available at \cite{MIRFLICKR}, with an additional 6K images from the Paris dataset available at \cite{ParisDataset}, as followed in \cite{Shi2014}; (\textit{ii}) \textsf{ImageNet}, a set of 1M images used in the ImageNet Object Localization Challenge, available at \cite{ImageNetKaggle}; (\textit{iii}) \textsf{IMDB-Wiki}, a set of all 500K images from the IMDB-Wiki face image dataset from \cite{IMDB-Wiki}; (\textit{iv}) \textsf{InstaCities}, a set of 1M images of 10 large cities \cite{InstaCities}. Although, we have worked with image datasets, our method is equally applicable to any other modalities such as speech, genomics, etc. 

In the following, we denote a query image by $\boldsymbol{I_q}$. For every dataset containing images $\{\boldsymbol{I_i}\}_{i=1}^N$, feature vectors $\{\boldsymbol{f_i}\}_{i=1}^N$ were extracted as follows: First, each image $\boldsymbol{I_i}$ was passed through the popular VGGNet \cite{Simonyan2015} (as its features are known to be useful for building good perceptual metrics \cite{Zhang2018}) to obtain intermediate feature vector $\boldsymbol{f'_i}$. Second, a softmax-based feature vector $\boldsymbol{f_i}$ of dimension $d = 1000$, corresponding to the 1000 classes in ImageNet, was extracted further from the VGGNet output vector $\boldsymbol{f'_i}$. In our experiments, we observed that the softmax-based features yielded superior retrieval recall (defined as \# points \emph{retrieved} belonging to the same class as the query point / \# points of the same class as the query point) as compared to the baseline VGGNet features, with exhaustive search on different datasets (see \cite[Table 1]{suppmat} for more details), even on those datasets which are different from ImageNet. Higher recall is particularly beneficial for NN search application in scenarios such as image plagiarism detection. The feature vectors were all non-negative and were subsequently unit-normalized. For every dataset, NN search was carried out using different methods (see below) on the same unit-normalized feature vectors extracted from 10,000 query images. That is, the aim was to efficiently identify those indices $i \in \{1,2,...,N\}$ for which the dot products between $\boldsymbol{q}$ (the feature vector of query image $\boldsymbol{I_q}$) and $\boldsymbol{f_i}$ (both feature vectors unit-normalized) exceeded some specified threshold $\rho$. 
For all datasets, the intrinsic dimensionality was computed using the number of singular values of the $d \times N$ data matrix that accounted for 99\% of its Frobenius norm. The intrinsic dimensionality for all datasets was more than 200 (for $d = 1000$).

Our binary splitting approach, using pool creation with summation (\textsc{Our-Sum}) and element-wise maximum (\textsc{Our-Max}), was compared to the following recent and popular techniques: (\textit{i}) Exhaustive search (\textsc{Exh}) through the entire database, for every query image;  (\textit{ii}) The recent GT-based technique from \cite{Engels2021}, known as \textsc{Flinng} (Filters to Identify Near-Neighbor Groups), using code from \cite{flinng}; (\textit{iii}) The popular randomized KD-tree and K-means based technique \textsc{Flann} for approximate NN search \cite{Muja2014}, using code from \cite{FlannPy} with default parameters; (\textit{iv}) The \textsc{Falconn} technique from \cite{Andoni2015} which uses multi-probe LSH, using the code from \cite{Falconn_imp}; (\textit{v}) \textsc{Falconn++} \cite{pham2022}, an enhanced version of \textsc{Falconn} which filters out potentially distant points from any hash bucket before querying using code from \cite{falcon_plus_plus}; (\textit{vi}) \textsc{Ivf}, the speedy inverted multi-index technique implemented in the highly popular \textsc{Faiss} library from Meta/Facebook \cite{Faiss,Babenko2014}; (\textit{vii}) \textsc{Hnsw}, a graph-based technique also used in \cite{Faiss};  
(\textit{viii}) The \textsc{Scann} library \cite{scann,Guo2020} which implements an anisotropic vector quantization (KMeans) algorithm. Our methods have no hyperparameters, but for all competing methods, results are reported by tuning hyperparameters so as to maximize the recall (defined below) -- see \cite[Sec. 5]{suppmat} for more details.

We did not compare with the GT-based techniques from \cite{Shi2014} and \cite{Iscen2018} as we could not find full publicly available implementations for these. Moreover, the technique from \cite{Engels2021} is claimed to outperform those in \cite{Shi2014} and \cite{Iscen2018}. We did implement the algorithm from \cite{Shi2014}, but it yielded significantly large query times. This may be due to subtle implementation differences in the way image lists are re-ranked in each of the $R$ iterations of their algorithm (see Sec.~\ref{sec:comparison}). In any case, our method produces 100\% recall and precision, unlike \cite{Shi2014,Iscen2018}.

\noindent\textbf{Non-streaming Setting:} The different methods are compared in terms of Mean Query Time (\texttt{MQT}), Mean Precision (\texttt{MP}) and Mean Recall (\texttt{MR}) for 10K queries as reported in Table \ref{tab:query_recall_comparison}. 
The similarity threshold ($\rho$) chosen for this setting is $0.8$. The recall is defined as as $\textrm{\#true positives}/(\textrm{\#true positives} + \textrm{\#false negatives})$. The precision is defined as $\textrm{\# true positives}/(\textrm{\#true positives} + \textrm{\# false positives})$. For each dataset, we have also mentioned the average number of neighbours ($\hat{k}$) having similarity greater than $\rho = 0.8$. The cumulative sums in Eqn.~\ref{eq:cumsums} for \textsc{Our-Sum} and the cumulative element-wise maximums for \textsc{Our-Max} were computed offline. The query implementation was done in C++ with the highest level of optimization in the g++ compiler. The codes for \textsc{Flinng}, \textsc{Falconn++}, \textsc{Scann} and \textsc{Hnsw} operate in the KNN search mode, as opposed to a range-based query for \textsc{Our-Sum}, \textsc{Our-Max}, \textsc{Falconn} and \textsc{Ivf}, for which we set $\rho = 0.8$. The \textsc{Flann} method has code support for both KNN and range-based queries, which we refer to as \textsc{Flann-K} and \textsc{Flann-R} respectively. For all KNN-based methods, we gave an appropriate $K$ as input, such that $K$ ground truth neighbors satisfy the similarity threshold of $\rho$. The time taken to compute $K$ was \emph{not} included in the query times reported for the KNN-based methods. All query times are reported for an Intel(R) Xeon(R) server with CPU speed 2.10GHz and 128 GB RAM. For all methods, the computed query times did not count the time taken to compute the feature vector for the query image.  

\noindent\textbf{Streaming Setting:} We now consider experiments in a streaming setting where new vectors are added to the database on the fly, i.e. the database is modified incrementally. In our experimental setting, 80\% of the full database was initially available to all algorithms for index creation. New points were incrementally added to the database, and after addition of 100 new points, a search query was fired. This process was carried out until the entire database was loaded into memory. 
The query images were created by introducing controlled perturbations such as dither and blur to images existing in the database (refer to \cite{suppmat} for more details). The similarity threshold for this setting is chosen to be $\rho=0.9$. Note that this setting is analogous to a typical image plagiarism detection setting, wherein (a) the database is updated with insertion of new points (without deletion) and requires fetching high similarity vectors, and (b) the submitted images could be subtly manipulated to escape plagiarism detection. High recall is desirable in such applications. 

The only update required to \textsc{Our-Sum} for insertion of $n_s$ new points will be to produce $n_s$ extra cumulative sums as in Eqn.~\ref{eq:cumsums} for a cost of $O(dn_s)$.  
For comparison, we have selected only those algorithms that include methods for the addition of new points without necessitating the rebuilding of the entire index from scratch. These include \textsc{Exh}, \textsc{Ivf}, \textsc{Hnsw} and \textsc{Flann-R}. Table~\ref{tab:streaming} reports the initial Index Build Time (\texttt{IBT}) for 80\% of the database, Mean Insertion Time (\texttt{MIT}) for new points, besides \texttt{MQT}, \texttt{MP} and \texttt{MR}.

\noindent \textbf{Discussion of Results on Static Datasets (Table~\ref{tab:query_recall_comparison}):} We observe that \textsc{Our-Sum} yields lower \texttt{MQT} on most datasets as compared to most other techniques, and that too with a guarantee of 100\% precision and recall. \textsc{Falconn} produces higher query times than \textsc{Our-Sum} or \textsc{Our-Max} and at the loss of some recall. \textsc{Falcon++} produces 100\% precision and recall but with much higher query times than \textsc{Our-Sum} or \textsc{Our-Max}. \textsc{Flinng} and \textsc{Flann-K} are efficient, but have significantly lower precision and recall than \textsc{Our-Sum} with many of the retrieved dot products being significantly lower than the chosen $\rho$. \textsc{Flann-R} yields very low recall despite its excellent precision and query time. \textsc{Hnsw} produces higher query times in cases where the number of neighbors is large with some loss of recall. The closest competitors to \textsc{Our-Sum} are \textsc{Scann} and \textsc{Ivf}, but they do not guarantee 100\% precision and recall. In addition, we note that as reported in \cite[Fig. 3a]{Guo2020}, the accuracy of the dot-products retrieved by \textsc{Scann} will depend on the chosen bit-rate, i.e. the chosen \emph{number of cluster centers} in KMeans, whereas \textsc{Our-Sum} has no such dependencies or parameter choice. This is a major conceptual advantage of \textsc{Our-Sum} over \textsc{Scann} (also see the `Streaming' setting below). As compared to exhaustive search, \textsc{Our-Sum} produces a speed gain between 4 to 120 depending on the dataset. \textsc{Our-Sum} outperforms \textsc{Our-Max} in terms of query time, but it is important note that \textsc{Our-Max} is capable of handling negative-valued descriptors, although we have not experimented with it in this paper. 
\renewcommand{\arraystretch}{1.5}
\begin{table}[h!]
\centering
\fontsize{6pt}{6pt}\selectfont
\begin{tabular}{|c|c|c|c|c|c|c||c|c|c|c|c||c|} 
 \hline
 \rowcolor{Gray}
 & Db. & \textsc{Our-S} & \textsc{Our-M} & \textsc{Ivf} & \textsc{Falc} & \textsc{Flann-R} & \textsc{Falc++} & \textsc{Flinng} & \textsc{Flann-K} & \textsc{Scann} & \textsc{Hnsw} & \textsc{Exh} \\\hline 
 \texttt{MQT} &  \textsf{MIRFL.} & 54.0 & 81 & 94 & 302 & 1 & 509 & 62 & 54 & 23 & 66 & 1222\\
 \texttt{MP} & \textsf{($\hat{k}\approx 1.7K$)} & 1.0 & 1.0 & 1.0 & $\approx$1.0 & 1.0 & 1.0 & 0.268 & 0.743 & 0.996 & 0.959 & 1.0 \\
\texttt{MR} & \textsf{} & 1.0 & 1.0 & 0.997 & 0.954& 0.323 & 1.0 & 0.268 & 0.743 & 0.996 & 0.959 & 1.0 \\\hline
\texttt{MQT} &  \textsf{ImgN.} & 4.9 & 7 & 103 & 179& 2 & 385 & 119 & 40 & 23 & 9 & 1276 \\
\texttt{MP} &  \textsf{($\hat{k} \approx 0.9K$)} & 1.0 & 1.0 & 1.0 & 1.0& 1.0 & 1.0 & 0.372 & 0.310 & 0.998 & 0.965 & 1.0 \\
\texttt{MR} & \textsf{} & 1.0 & 1.0 & 0.997 & 0.958& 0.217 & 1.0 & 0.372 & 0.310 & 0.998 & 0.965 & 1.0\\\hline
\texttt{MQT} &  \textsf{IMDB.} & 98.8 & 249 & 49 & 415& 1.0 & 739 & 46 & 98 & 60 & 1076 & 800\\
\texttt{MP} &  \textsf{($\hat{k}\approx 7K$)} & 1.0 & 1.0 & 1.0 & $\approx$1.0 & 1.0 & 1.0 & 0.351 & 0.656 & 0.997 & 0.985& 1.0\\
\texttt{MR} & \textsf{} & 1.0 & 1.0 & 0.995 & 0.995 & 0.191 & 1.0 & 0.351 & 0.656& 0.997 & 0.985 & 1.0 \\\hline
\texttt{MQT} & \textsf{InstaC.}  & 58.0 & 110 & 90 & 330 & 1 & 601 & 57 & 61 & 25 & 298 & 1174 \\
\texttt{MP} & \textsf{($\hat{k}\approx 3K$)}  & 1.0 & 1.0 & 1.0 & $\approx$1.0& $\approx$1.0 & 1.0 & 0.293 & 0.745& 0.996 & 0.952 & 1.0 \\
\texttt{MR} & \textsf{} & 1.0 & 1.0 & 0.990 & 0.964& 0.324 & 1.0 & 0.293 & 0.745& 0.996 & 0.952& 1.0 \\
 \hline
 \end{tabular}
\vspace{2mm}
\caption{Comparison of mean query time (\texttt{MQT}, $\downarrow$) in milliseconds, and mean precision (\texttt{MP}, $\uparrow$) and mean recall (\texttt{MR}, $\uparrow$) values for our method with summation pooling (\textsc{Our-S}.); our method with max pooling (\textsc{Our-M}.); \textsc{Ivf} from the \textsc{Faiss} package \cite{Faiss,Babenko2014}; \textsc{Hnsw}  from the \textsc{Faiss} package \cite{Faiss,Malkov2018}; \textsc{Falconn} \cite{Andoni2015,Falconn_imp}; \textsc{Falconn+} \cite{pham2022}; \textsc{Flinng}\cite{Engels2021}; \textsc{Flann}\cite{Muja2014} with KNN-based (\textsc{Flann-K}) and range-based queries (\textsc{Flann-R}); \textsc{Scann} \cite{scann} and exhaustive search (\textsc{Exh}). For all methods $\rho = 0.8$, see text for more details. $\hat{k} \triangleq$ avg. number of database vectors satisfying $\boldsymbol{q}^t \boldsymbol{f_i} \geq \rho$.} 
\vspace{-0.7cm}
\label{tab:query_recall_comparison}
\end{table}
\begin{table}[h!]
\fontsize{7pt}{7pt}\selectfont
\begin{subtable}[h]{0.45\textwidth}
\begin{tabular}{|c|c|c|c|c|}
\hline\rowcolor{Gray}
\multicolumn{5}{|c|}{Index Build Time (in ms) $\downarrow$}
\\\hline\rowcolor{LightGray}
Alg.$\downarrow$, Db. $\rightarrow$ & \textsf{ImgN.} & \textsf{IMDB.} & \textsf{MIRFL.} & \textsf{InstaC.} \\
\hline        
\textsc{Our-Sum} & 3789 & 2000 & 3774 & 3765 \\\hline
\textsc{Exh.} &  NA & NA & NA & NA\\\hline
\textsc{Ivf} & 15000 & 9960 & 15410 & 18722\\\hline
\textsc{Hnsw} & 7e5 & 2e5 & 4e5 & 4e5 \\\hline
\textsc{Flann-R} & 2e6 & 9e5 & 2e6 & 2e6 \\\hline
\end{tabular}
\end{subtable}
\hspace{1cm}
\begin{subtable}[h]{0.45\textwidth}
\begin{tabular}{|c|c|c|c|c|}
\hline\rowcolor{Gray}
\multicolumn{5}{|c|}{Average Insertion Time (in ms) $\downarrow$}
\\\hline\rowcolor{LightGray}
Alg.$\downarrow$, Db. $\rightarrow$ & \textsf{ImgN.} & \textsf{IMDB.} & \textsf{MIRFL.} & \textsf{InstaC.} \\
\hline        
\textsc{Our-Sum} & 0.28 & 0.29 & 0.28 & 0.29\\\hline
\textsc{Exh.} &  $\sim 0$ & $\sim 0$ & $\sim 0$ & $\sim 0$\\\hline
\textsc{Ivf} & 2.2 & 2.7 & 2.2 & 2.7\\\hline
\textsc{Hnsw} & 236 & 239 & 222 & 219 \\\hline
\textsc{Flann-R} & 42 & 53 & 27 & 26 \\\hline
\end{tabular}
\end{subtable}\\ \\ \\

\begin{subtable}[h]{0.45\textwidth}
\begin{tabular}{|c|c|c|c|c|}
\hline\rowcolor{Gray}\rowcolor{LightGray}
\multicolumn{5}{|c|}{Mean Query Time (in ms) $\downarrow$}
\\\hline\rowcolor{LightGray}
Alg.$\downarrow$, Db. $\rightarrow$ & \textsf{ImgN.} & \textsf{IMDB.} & \textsf{MIRFL.} & \textsf{InstaC.} \\
\hline        
\textsc{Our-Sum} & 18 & 121 & 86 & 115\\\hline
\textsc{Exh.} &  609 & 315 & 607 & 605\\\hline
\textsc{Ivf} & 183 & 67 & 145 & 132\\\hline
\textsc{Hnsw} & 10 & 168 & 13 & 70\\\hline
\textsc{Flann-R} & 6 & 4 & 5 & 5\\\hline
\end{tabular}
\end{subtable}
\hspace{-0.2cm}
\begin{subtable}[h]{0.45\textwidth}
\begin{tabular}{|c|c|c|c|c|}
\hline\rowcolor{Gray}\rowcolor{LightGray}
\multicolumn{5}{|c|}{Mean Precision $\uparrow$, Mean Recall $\uparrow$}
\\\hline\rowcolor{LightGray}
Alg.$\downarrow$, Db. $\rightarrow$ & \textsf{ImgN.} & \textsf{IMDB.} & \textsf{MIRFL.} & \textsf{InstaC.} \\
\hline        
\textsc{Our-Sum} & 1,1 & 1,1 & 1,1 & 1,1\\\hline
\textsc{Exh.} &  1,1 & 1,1 & 1,1 & 1,1\\\hline
\textsc{Ivf} & 1,0.99 & 1,0.99  & 1,0.99 & 1,0.99\\\hline
\textsc{Hnsw} & 0.96,0.96 & 0.98,0.98 & 0.97,0.97 & 0.97,0.97\\\hline
\textsc{Flann-R} & 1,0.31 & 1,0.59 & 1,0.49 & 1,0.54\\\hline
\end{tabular}
\end{subtable}
\vspace{2mm}
\caption{Streaming results: Index Build Time (\texttt{IBT}) for 80\% of the database, mean insertion time (\texttt{MIT}) for new points, \texttt{MQT}, \texttt{MP} and \texttt{MR}. For \textsf{IMDB}, the number of insertion operations and search queries is $\sim$700 and $\sim$1350 respectively. For all other datasets, the numbers are $\sim$1400 and $\sim$2600 respectively.}
\vspace{-0.9cm}
\label{tab:streaming}
\end{table}

\noindent\textbf{Discussion of Results on Datasets with Streaming (Table \ref{tab:streaming})}: 
It is clear that the Index Build Time (\texttt{IBT}) and Mean Insertion Time (\texttt{MIT}) of our method clearly surpasses that of all other methods across all datasets. Furthermore, the space taken for storing the index is only $O(Nd)$ in our method. Since \textsc{Our-Sum} only has to compute cumulative sums for the newly inserted points, the index update time is very low when compared with other methods. \textsc{Flann-R} consistently produces lower mean query times, but this comes at the cost of low recall and high insertion time, making it unusable for streaming applications. Similarly, \textsc{Hnsw} shows significant gains in \texttt{MQT} for \textsf{MIRFLICKR}, \textsf{ImageNet} and \textsf{InstaCities} datasets, but exhibits a large insertion time across all datasets. 

\noindent\textbf{Analysis of Pruning (Non-Streaming):} For each dataset, we also computed the number of pools that got rejected in every `\textbf{round}' of binary splitting (because the dot product with the query image fell below threshold $\rho$), beginning from round 0 to $\lceil \log_2 N \rceil$. (Referring to Alg.~\ref{alg:binsplt}, we see that the procedure implements an inorder traversal of the binary tree implicitly created during binary splitting. However the same tree can be traversed in a level-wise manner, and each level of the tree represents one `\textbf{round}' of binary splitting.) The corresponding histograms for $\rho \geq 0.7$ are plotted in Fig. 2 of \cite{suppmat}, for \textsc{Our-Sum}. As can be seen, a large number of rejections occur from round 14 onward in \textsf{MIRFLICKR} and \textsf{InstaCities}, from round 11 onward in \textsf{ImageNet}, and from round 16 onward in \textsf{IMDB-Wiki}. As a result, the \texttt{MQT} is lowest for \textsf{ImageNet} than for other datasets, and also much lower for \textsf{MIRFLICKR} and \textsf{InstaCities} than for \textsf{IMDB-Wiki}. A similar set of histograms are plotted for \textsc{Our-Max} in Fig. 3 of \cite{suppmat}. Furthermore, we plotted histograms of dot products between every query feature vector and every gallery feature vector, across 10,000 queries for all datasets. These are plotted in Fig.~\ref{fig:exponential}, clearly indicating that the vast majority of dot products lie in the bin $(0,0.1]$ or $(0.1,0.2]$ (also see Fig. 1 of \cite{suppmat}, which shows negative logarithms of the histograms for clearer visualization of values close to 0). This supports the hypothesis that most items in $\mathcal{D}$ are dissimilar to a query vector. However, as seen in Fig.~\ref{fig:exponential}, the percentage of dot products in the $(0,0.1]$ bin is much larger for \textsf{ImageNet}, \textsf{MIRFLICKR} and \textsf{InstaCities} than for \textsf{IMDB-Wiki}. This tallies with the significantly larger speedup obtained by binary splitting (both \textsc{Our-Sum} and \textsc{Our-Max}) for \textsf{ImageNet}, \textsf{MIRFLICKR} and \textsf{InstaCities} as compared to \textsf{IMDB-Wiki} -- see Table~\ref{tab:query_recall_comparison}. 

\section{Theoretical Analysis}
\label{sec:theoretical_analysis}
\begin{figure}
    \centering
    \includegraphics[scale=0.15]{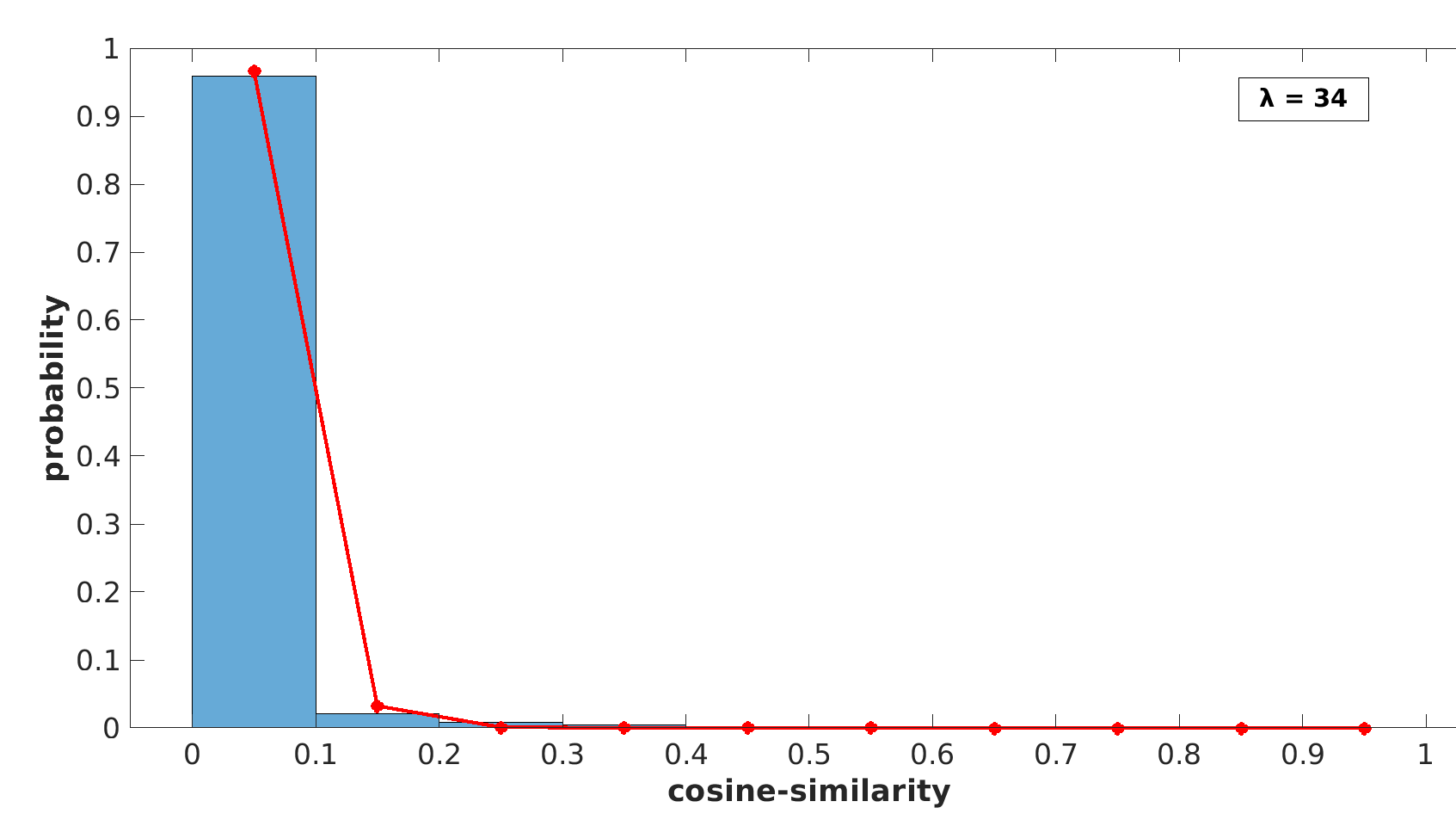}
    \includegraphics[scale=0.15]{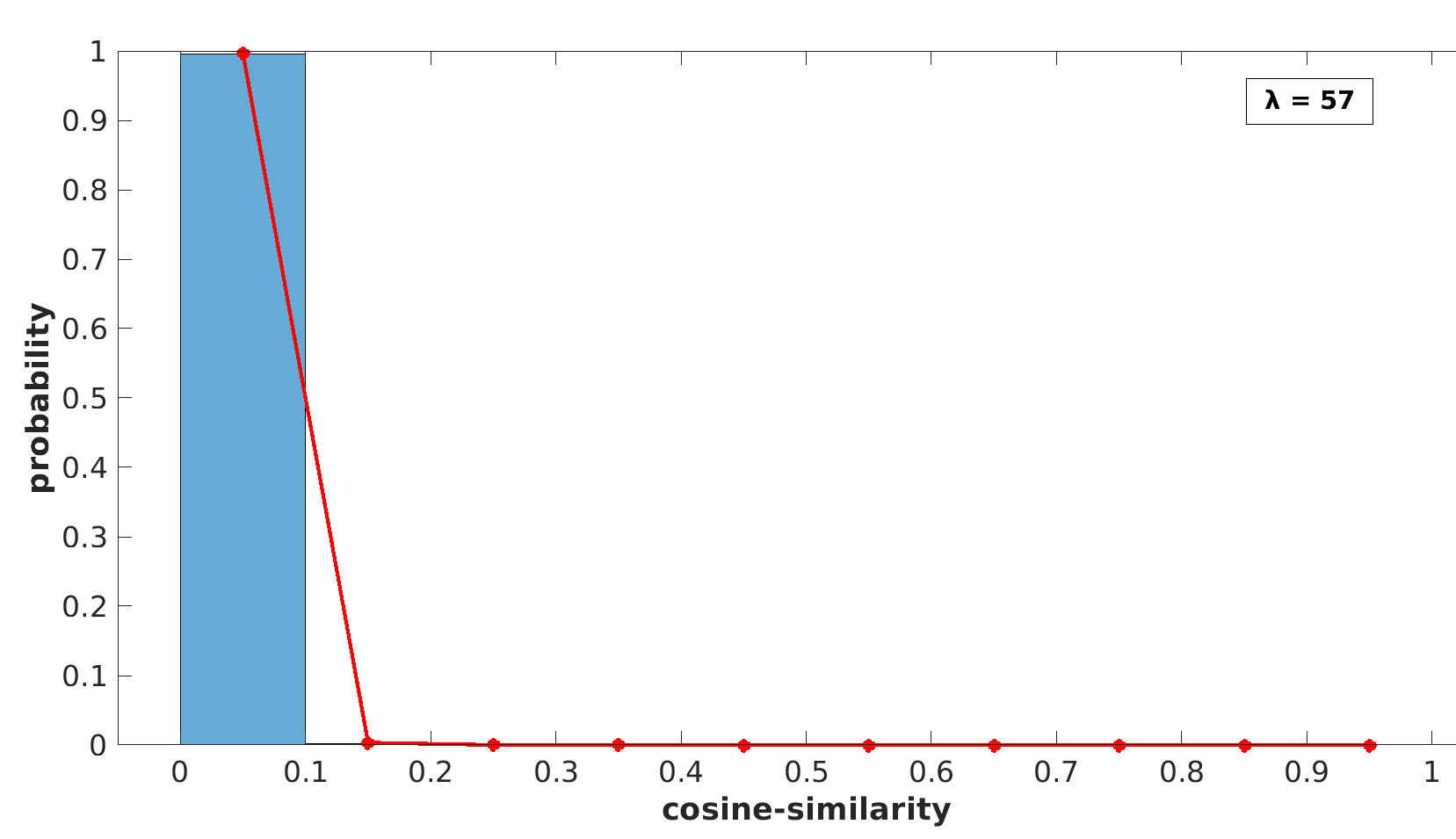}
    \includegraphics[scale=0.15]{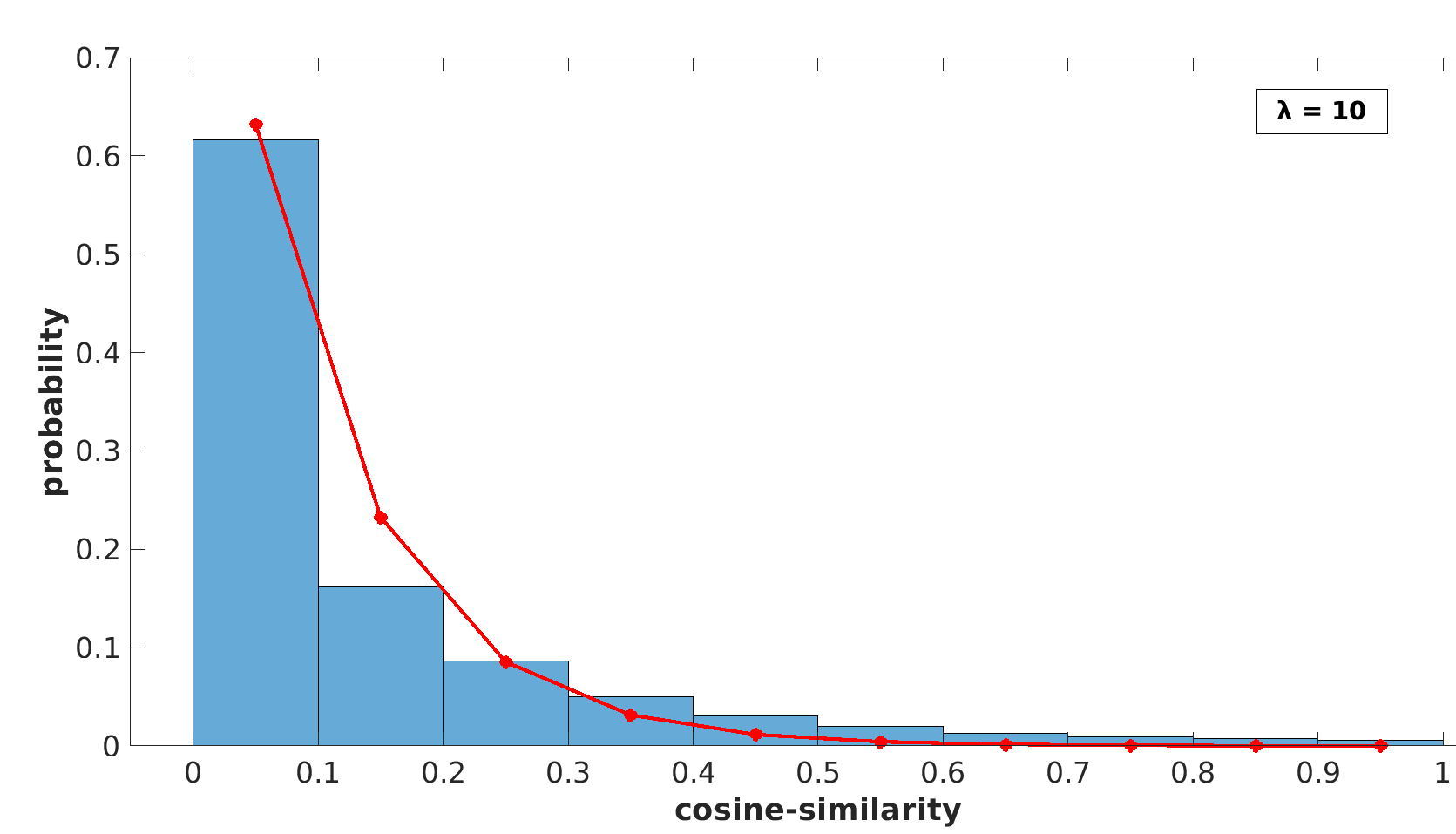}
    \includegraphics[scale=0.15]{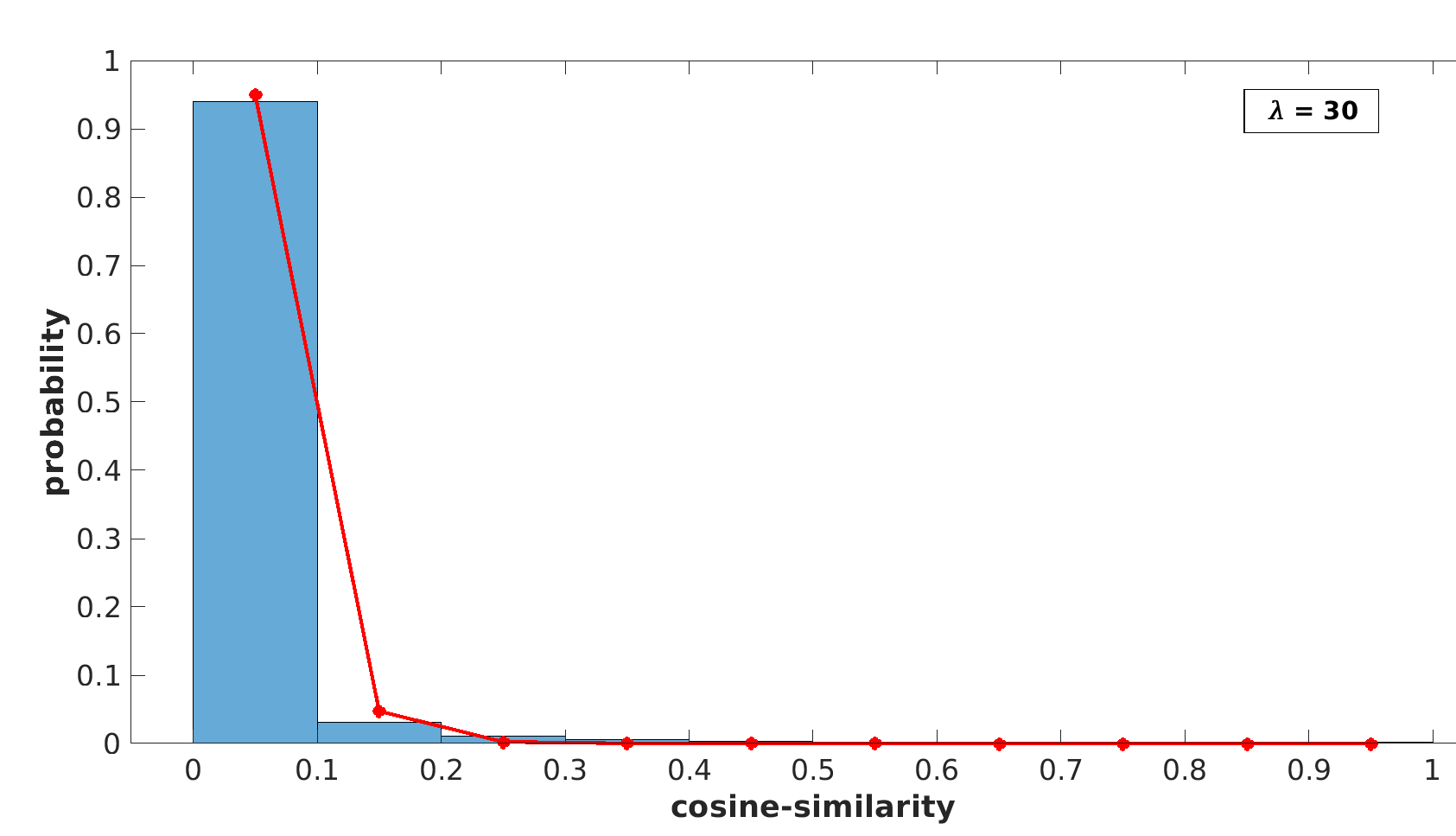}
    \caption{Normalized histograms (bar graphs in blue), and best-fit exponential distribution approximations (red curves), of dot products between feature vectors of query images (10,000 in number) and all images of \textsf{MIRFLICKR}, \textsf{ImageNet}, \textsf{IMDB-Wiki} and \textsf{InstaCities} databases (left to right, top to bottom). Also see Fig. 1 of \cite{suppmat} for a plot of negative log of the histograms for clearer visualization of the smaller probability values.}
    \label{fig:exponential}
    \vspace{-0.001cm}
    
\end{figure}

For the problem of binary GT, different variants of the binary splitting approach have been shown to require just $s \log_2 N + O(s)$ or $s \log_2 (N/s) + O(s)$ tests, depending on some implementation details \cite[Theorems 1.2 and 1.3]{Aldridge2019}. Here $s$ stands for the number of defectives out of $N$. However in binary GT, a positive result on a pool necessarily implies that at least one of the participants was defective. In our application for NN search, a pool $\boldsymbol{y}$ can produce $\boldsymbol{q}^t \boldsymbol{y} \geq \rho$ for query vector $\boldsymbol{q}$, even if \emph{none} of the vectors that contributed to $\boldsymbol{y}$ produce a dot product with $\boldsymbol{q}$ that exceeds or equals $\rho$. Due to this important difference, the analysis from \cite{Aldridge2019} does not apply in our situation. Furthermore, it is difficult to pin down a precise \emph{deterministic} number of tests in our technique. Instead, we resort to a distributional assumption on each dot product $\boldsymbol{q}^t \boldsymbol{f_i}$. We assume that these dot products are independently distributed as per a truncated normalized exponential (TNE) distribution:
\begin{equation}
p(x; \lambda) = 
\begin{cases}
\dfrac{\lambda e^{-\lambda x}}{1-e^{-\lambda}}, \textrm{ for } 0 \leq x \leq 1. \\
0, \textrm{ otherwise.}
\end{cases}
\label{eq:TNE}
\end{equation}
This distribution is obtained by truncating the well-known exponential distribution to the $[0,1]$ range and normalizing it so that it integrates to one. Larger the value of $\lambda$, the more rapid is the `decay' in the probability of dot product values from 0 towards 1. Our empirical studies show that for softmax features, this is a reasonable model for dot product values commonly encountered in queries in image retrieval, as can be seen in Fig.~\ref{fig:exponential} for all datasets we experimented with. Also, the best-fit $\lambda$ values for \textsf{MIRFLICKR}, \textsf{ImageNet}, \textsf{IMDB-Wiki} and \textsf{InstaCities} are respectively $34, 57, 10, 30$. The TNE model fit may not be perfect, but less than perfect adherence to the TNE model does not change the key message that will be conveyed in this section.

Our aim now is to determine the probability that a pool $\boldsymbol{y}$ with some $L$ participating vectors will produce a dot product value $\boldsymbol{q}^t \boldsymbol{y}$ below a threshold $\rho$. Note again that such a pool will get pruned away in binary splitting, and therefore we would like this probability to be high. Now, the $L$ individual dot-products (of $\boldsymbol{q}$ with vectors that contributed to $\boldsymbol{y}$) are independent and they are all distributed as per the TNE from Eqn.~\ref{eq:TNE}. The mean and variance of a TNE variable with parameter $\lambda$ are respectively given by 
$\left(1/\lambda + 1/(1-e^{\lambda})\right)$ and 
$\left(1/\lambda^2 - e^{\lambda}/(e^{\lambda}-1)^2\right)$. We now apply the central limit theorem (CLT), due to which we can conclude that $\boldsymbol{q}^t \boldsymbol{y}$ is (approximately) Gaussian distributed with a mean $\mu \triangleq L\left(1/\lambda + 1/(1-e^{\lambda})\right)$ and variance $\sigma^2 \triangleq L\left(1/\lambda^2 - e^{\lambda}/(e^{\lambda}-1)^2\right)$. With this, we can easily compute $P(\boldsymbol{q}^t \boldsymbol{y} \leq \rho)$ given a value of $L$ and $\rho$. Our numerical experiments, illustrated in Fig. 3 of \cite{suppmat}, reveal that this probability increases with $\lambda$ and decreases as $L$ increases. This tallies with intuition, and supports the hypothesis that a larger number of pools will be pruned away in a given round if there is more rapid decrease in the probability of dot product values from 0 to 1. 

We are now equipped to determine the expected number of tests (dot products) to be computed per query, given such a TNE distribution for the dot products. Given a value of $\rho$, let us denote the probability that a pool with $L$ participants will get pruned away, by $p_L(\rho)$. For \textsc{Our-Sum}, for any $L \geq 6$, the probability $p_L(\rho)$ is computed using the TNE and CLT based method described earlier. For $L < 6$, the Gaussian distribution can be replaced by a truncated Erlang distribution, as the sum of independent exponential random variables follows an Erlang distribution. Even in this case, $P(\boldsymbol{q}^t \boldsymbol{y} \leq \rho)$ increases with $\lambda$ and decreases with $L$. 

In round 1, there is only one pool and it contains all $N$ vectors. It will get pruned away with probability $p_N(\rho)$, for a given fixed threshold $\rho$ and fixed $\lambda$. The expected number of pools in round 2 will be $Q_2 \triangleq 2(1-p_N(\rho))$, each of which will be pruned away with probability $p_{N/2}(\rho)$. The factor 2 is because each pool gets split into two before sending it to the next round of binary splitting. The expected number of pools in round 3 will be $Q_3 \triangleq 2(1-p_{N/2}(\rho))Q_2$. Likewise, in the $k$th round, the expected number of pools is $Q_k = 2(1-p_{N/2^{k-2}}(\rho))Q_{k-1}$. Hence the expected number of tests per query in our approach will be given by:
\begin{equation}
E(\rho) \triangleq 1+\frac{1}{2}\sum\limits_{i=2}^{\lceil \log_2 N +1 \rceil} Q_i, 
\label{eq:E_rho}
\end{equation}
as there are at most $\lceil \log_2 N +1 \rceil$ rounds. The factor $\frac{1}{2}$ is due to the fact that the number of tests (i.e. dot product computations) is always half the number of pools (see second para. of Sec.~\ref{sec:method}).
The value of $E(\rho)$ decreases with $\lambda$ as $p_N(\rho), p_{N/2}(\rho),...,p_{N/2^{k-2}}(\rho)$ all increase with $\lambda$. A similar analysis for \textsc{Our-Max} is presented in \cite[Sec. 4]{suppmat}.
\begin{wraptable}{r}{5.5cm} 
\fontsize{6pt}{6pt}\selectfont
\begin{tabular}{ |c|c|c|c| }
 \hline
 \rowcolor{Gray}
Database & & &\\\hline
 \rowcolor{Gray}
\textsf{MIRFLICKR}, $N=1M$ & $\rho = 0.7$ & $\rho = 0.8$ & $\rho = 0.9$\\\hline
Empirical (Sum) & 51690 &	44194 &	37788 \\
Empirical (Max) & 75782 &	60752 &	47531 \\
Theoretical (Sum) &	63838 & 	57553	& 50094 \\
Theoretical U.B. (Max) &	371672 & 	270408	& 195140 \\ \hline
 \rowcolor{Gray}
\textsf{ImageNet}, $N = 1M$ & $\rho = 0.7$ & $\rho = 0.8$ & $\rho = 0.9$\\\hline
Empirical (Sum)&	13852	& 12339 & 	10801\\ 
Empirical (Max) & 6946 & 5713 &	4551 \\
Theoretical	(Sum)& 34524	& 32603	& 31345\\
Theoretical U.B. (Max) &	14160 & 	6267	& 2775 \\ \hline
 \rowcolor{Gray}
\textsf{IMDB-Wiki}, $N = 500K$ & $\rho = 0.7$ & $\rho = 0.8$ & $\rho = 0.9$\\\hline			
Empirical (Sum) &	160020 &	141290 &	125050\\
Empirical (Max) & 240947 &	197370 &	158284 \\
Theoretical (Sum)&	113600 &	99552	& 87835\\
Theoretical U.B. (Max) &	1426259 & 	1336543	& 1249448 \\ \hline
 \rowcolor{Gray}
\textsf{InstaCities}, $N = 1M$	& $\rho = 0.7$ & $\rho = 0.8$ & $\rho = 0.9$\\\hline
Empirical (Sum)&	72729 &	62741	& 54125\\
Empirical (Max) & 105122 &	85188 &	67441 \\
Theoretical (Sum)&	71021 & 	63453 &	57944\\
Theoretical U.B. (Max) &	566498 & 	445181	& 346927 \\ \hline
\end{tabular}
\caption{Empirically observed average and theoretically predicted expected number of dot products (Eqn.~\ref{eq:E_rho}) for queries with $\rho \in \{0.7,0.8,0.9\}$ for different datasets. See explanation at the end of Sec.~\ref{sec:theoretical_analysis}. Note that `Sum' and `Max' denote algorithms \textsc{Our-Sum} and \textsc{Our-Max} respectively.}
\vspace{-1.1cm}
\label{tab:emp_theor}
\end{wraptable}
Our theoretical analysis depends on $\lambda$, but note that our algorithm is completely agnostic to $\lambda$ and does not use it in any manner. For values of $\rho \in \{0.7,0.8,0.9\}$, we compared the theoretically predicted average number of dot products ($E(\rho)$) to the experimentally observed number for all datasets. These are presented in Tab.~\ref{tab:emp_theor}, where we observe that for \textsf{MIRFLICKR}, \textsf{ImageNet} and \textsf{InstaCities}, the theoretical number for \textsc{Our-Sum} is generally \emph{greater} than the empirical one. This is because the dot products for these datasets decay slightly faster than what is predicted by a TNE distribution, as can be seen in Fig.~\ref{fig:exponential}. Conversely, for \textsf{IMDB-Wiki}, the theoretical number for \textsc{Our-Sum} is \emph{less} than the empirical one because the dot products decay slower for this dataset than what is predicted by a TNE distribution. 
For \textsc{Our-Max}, the difference between empirical and theoretical values is larger, as explain in \cite{suppmat}.

\section{Conclusion}
\label{sec:conclusion}
We have presented a simple and intuitive multi-round group testing approach for range-based near neighbor search. Our method outperforms exhaustive search by a large factor in terms of query time without losing accuracy, on datasets where query vectors are dissimilar to the vast majority of vectors in the gallery. The larger the percentage of dissimilar vectors, the greater is the speedup our algorithm will produce. Our technique has no tunable parameters. It is also easily applicable to situations where new vectors are dynamically added to an existing gallery, without requiring any expensive updates. Apart from this, our technique also has advantages in terms of accuracy of query results (especially recall) over many recent, state of the art methods. A significant limitation of our method is that it currently only supports cosine distances and is applicable only to databases with a sharp decay in feature similarity  distribution, as in the case with softmax features. At present, a pool may produce a large dot product value with a query vector, even though many or even all the members of the pool are dissimilar to the query vector. Developing techniques to reduce the frequency of such situations, possibly using various LSH techniques, as well as testing on billion-scale datasets, are important avenues for future work. 
The implementation of our algorithms can be found at \url{https://github.com/Harsh-Sensei/GroupTestingNN}.

\noindent\textbf{Acknowledgement:} The authors thank the Amazon IIT Bombay AI-ML Initiative for support for the work in this paper.

\bibliographystyle{splncs04}
\bibliography{main}

\clearpage

\begin{center}
\section*{\Huge \textsc{Supplementary Material}}
\end{center}

\section{Contents}
This supplemental material contains the following:
\begin{enumerate}
\item A figure showing the negative logarithm of the normalized histograms of dot products between feature vectors of query images and those of all gallery images from each of the four datasets we experimented with in the main paper. These are shown in Fig.~\ref{fig:histograms_dotproducts}. Compare with Fig. 2 of the main paper where normalized histograms are shown.
\item A comparison of the techniques \textsc{Our-Sum} and \textsc{Our-Max} proposed in this paper to recent group testing based NN search methods such as \cite{Engels2021,Shi2014,Iscen2018} is presented in Sec.~\ref{sec:comparison_gt}.
\item A discussion regarding the limitation of using the recent advances in the compressed sensing literature to the NN search problem is presented in Sec.~\ref{sec:CS_limitations}.
\item Statistics for the average number of points pruned away in each round of binary splitting in our method on each of the four databases, are presented in Fig.~\ref{fig:rejection_round_sum} for \textsc{Our-Sum} and Fig.~\ref{fig:rejection_round_max} for \textsc{Our-Max}. Notice that for both the proposed algorithms (\textsc{Our-Sum} and \textsc{Our-Max}) there is nearly no pruning encountered during the initial rounds. But as the adaptive tests continue, negative tests are encountered in further rounds, and hence pools get pruned away. Surprisingly, most negative tests for \textsc{Our-Max} are encountered in the last round in contrast to that for \textsc{Our-Sum}.

\item Our numerical experiments, illustrated in Fig.~\ref{fig:prob_dotproducts} reveal that $P(\boldsymbol{q}^t \boldsymbol{y} \leq \rho)$ increases with $\lambda$ and decreases as $L$ increases. This tallies with intuition, and supports the hypothesis that a larger number of pools will be pruned away in a given round if there is more rapid decrease in the probability of dot product values from 0 to 1. 

\item A theoretical analysis of the \textsc{Our-Max} algorithm is presented in Sec.~\ref{sec:theory_our_max}.

\item The hyper-parameters for various competing algorithms are presented in Sec.~\ref{sec:hyperparameter}.
\item Details of augmentation to images for generating queries in a streaming setting are presented in Sec.~\ref{sec:image_aug} and Fig.~\ref{fig:augmentation}.
\item A comparison of softmax features (on top of VGGNet features) versus VGGNet features in terms of retrieval precision and retrieval recall is presented in Table~\ref{tab:softmax_VGGNet} and Sec.~\ref{sec:comp_vgg16}. 
\item Experiments with a low similarity value, i.e. $\rho$ (=0.3), and experiments on a dataset with 12-million points are presented in Sec. ~\ref{sec:other_exp}.
\end{enumerate}

\begin{figure}[H]
\centering
\includegraphics[scale=0.22]{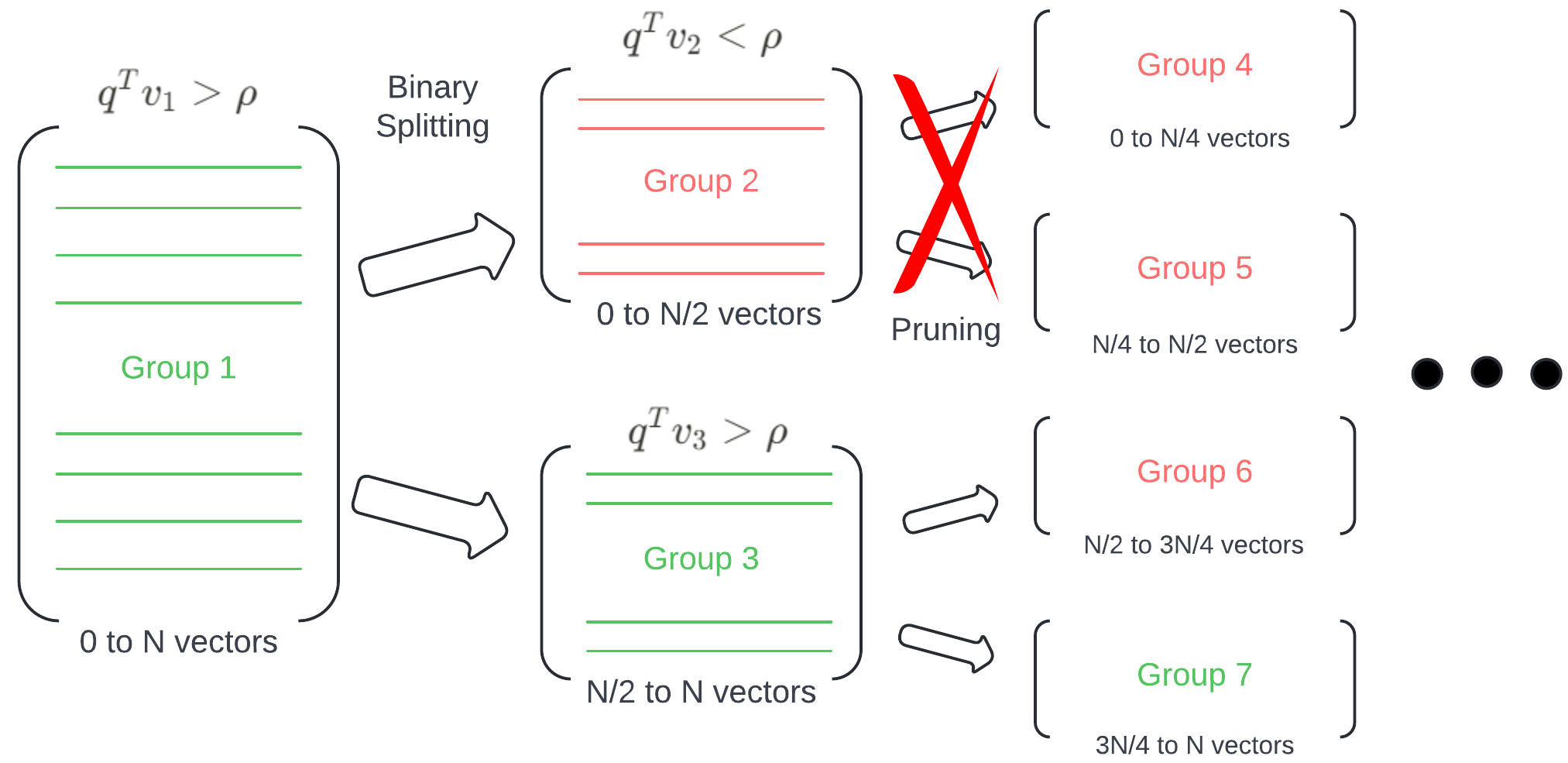}  
\caption{Schematic of our GT algorithm for NN search}
\label{fig:schematic}
\end{figure}

\begin{figure*}
    \centering
    \includegraphics[scale=0.17]{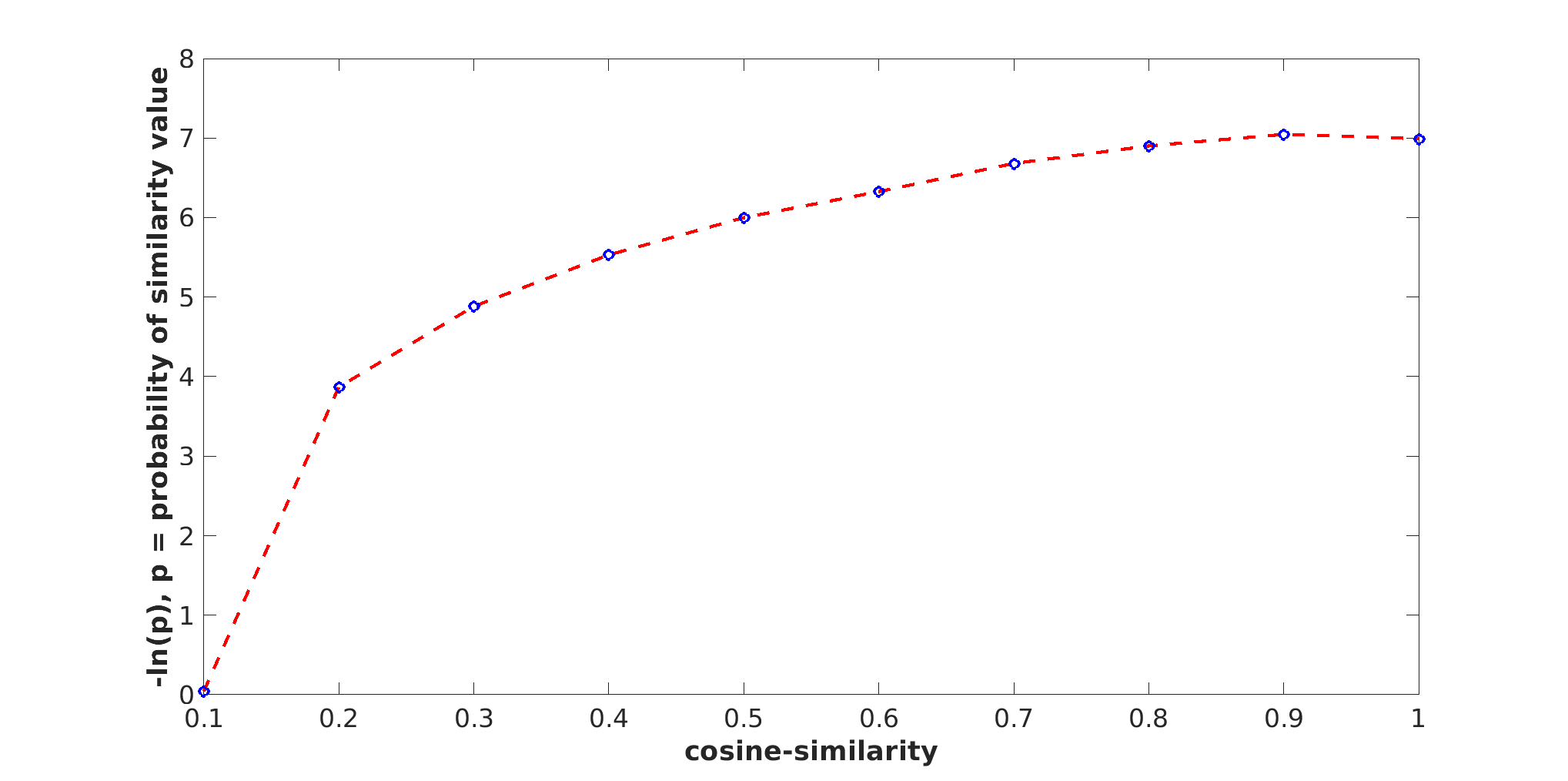}
    \includegraphics[scale=0.17]{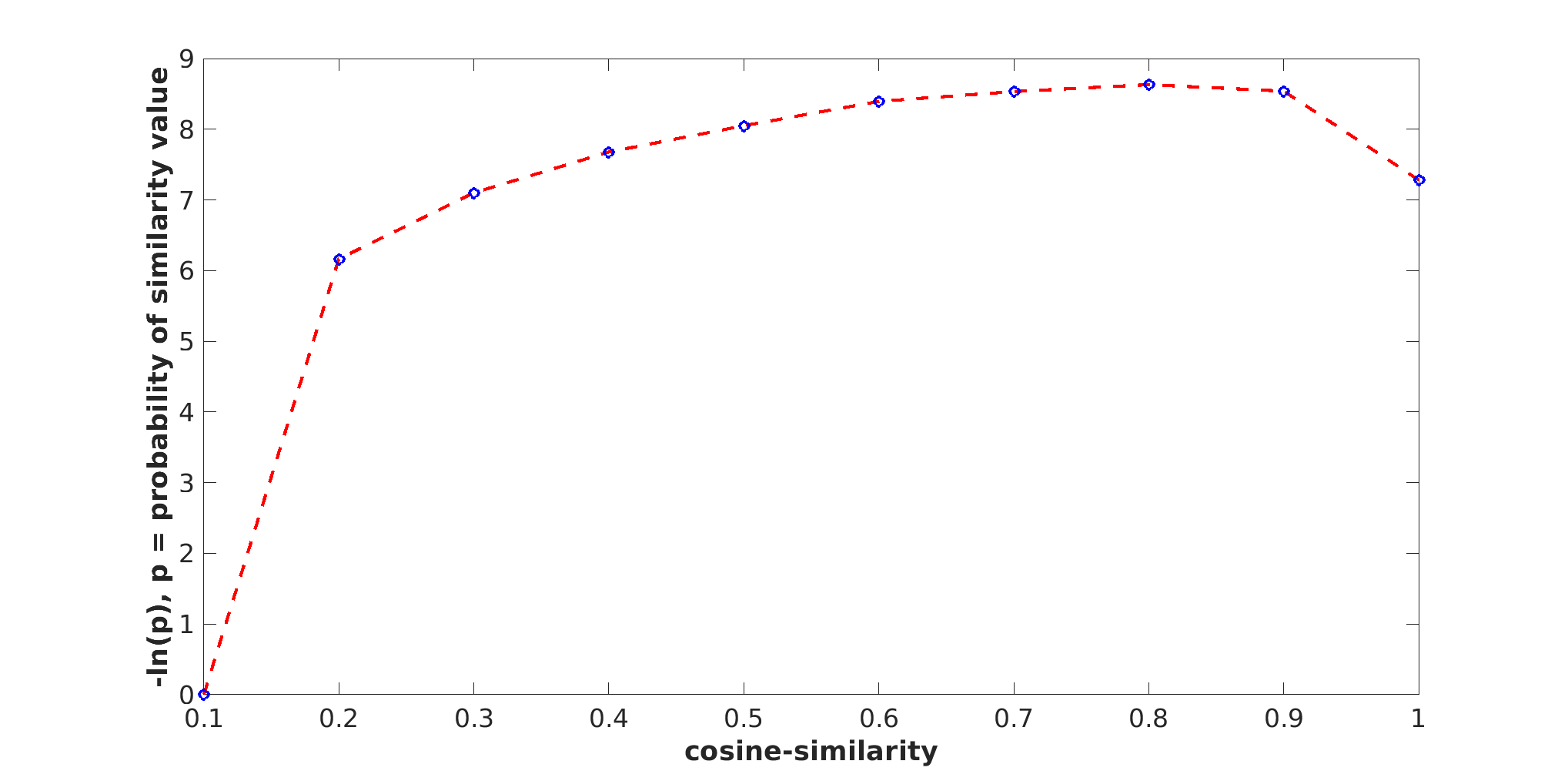}
    \includegraphics[scale=0.17]{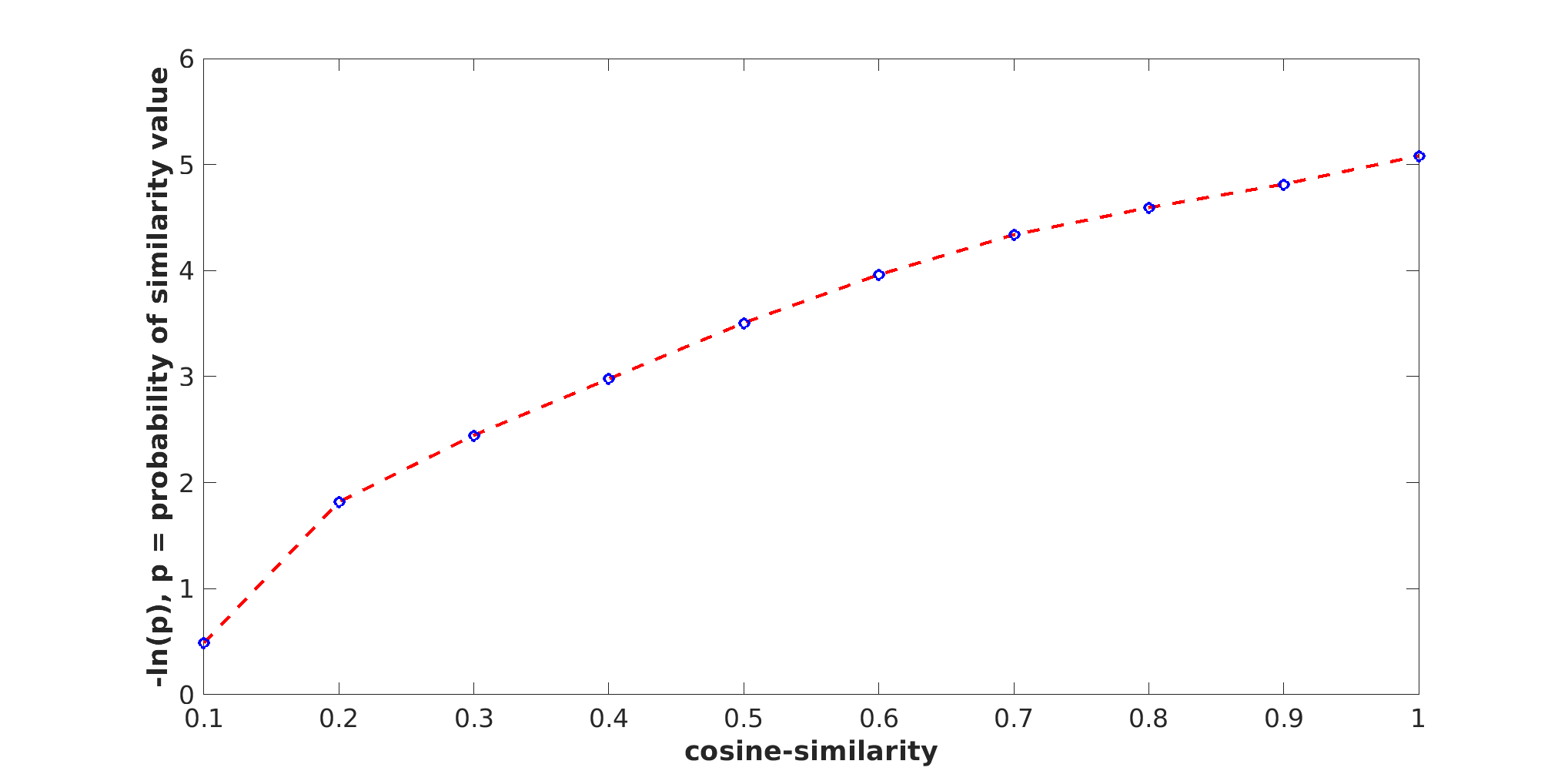}
    \includegraphics[scale=0.17]{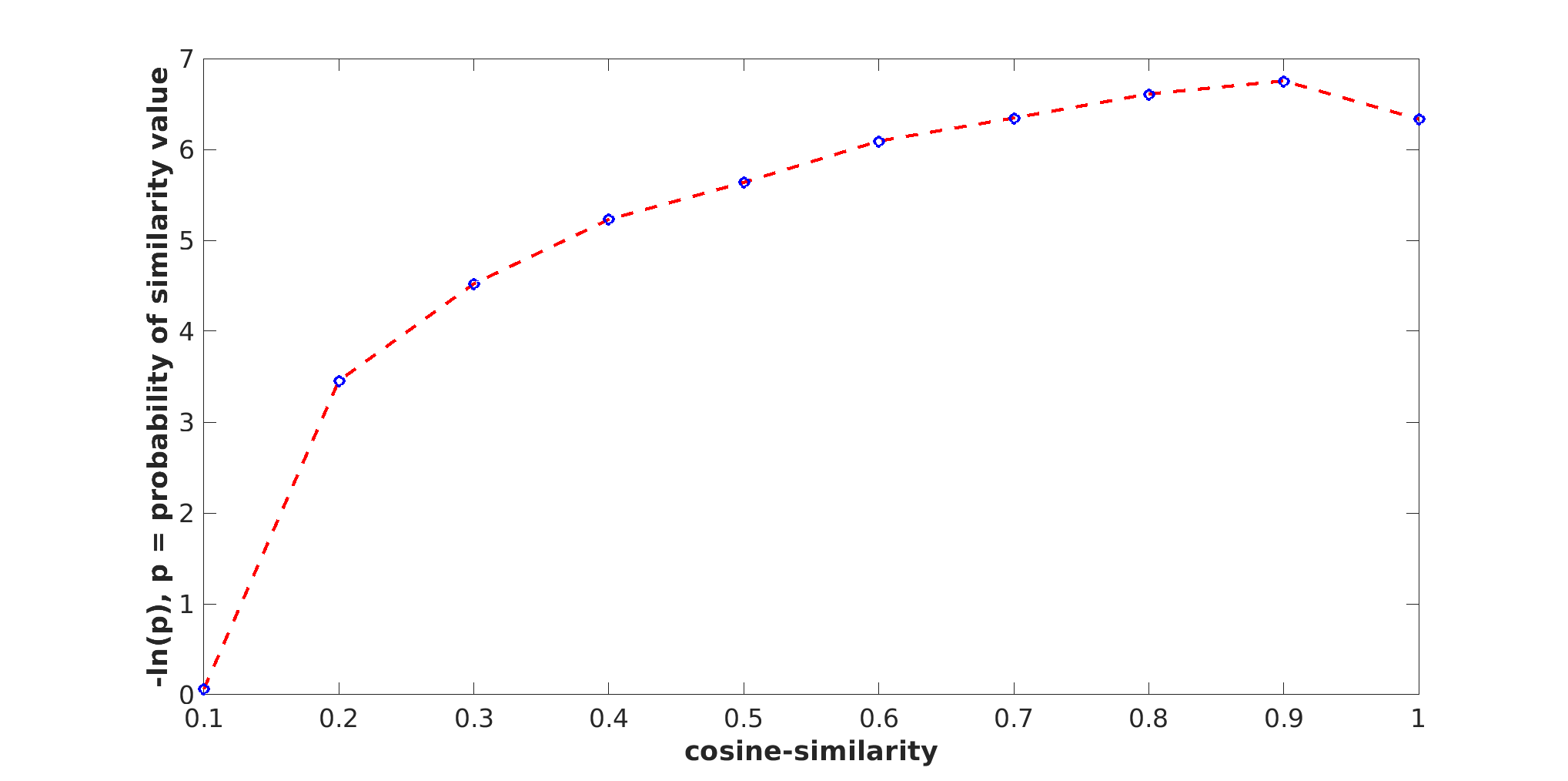}
    \caption{Negative logarithm of normalized histograms of dot products between feature vectors of query images (10K in number) and those of all gallery images of \textsf{MIRFLICKR}, \textsf{ImageNet}, \textsf{IMDB-Wiki} and \textsf{InstaCities} (left to right, top to bottom). Compare to Fig. 2 of the main paper.}
    \label{fig:histograms_dotproducts}
\end{figure*}
\begin{figure*}
    \centering
    \includegraphics[scale=0.17]{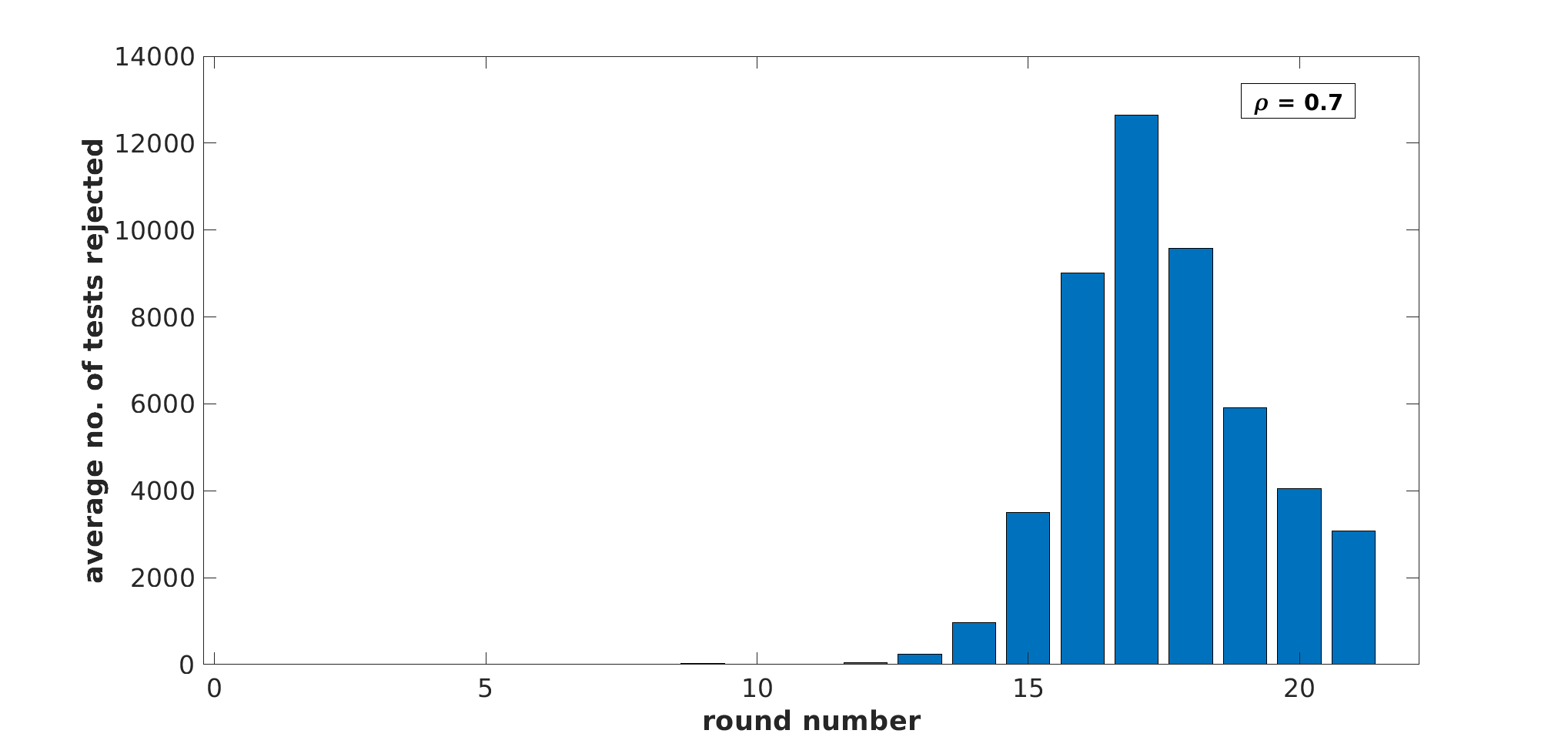}
    \includegraphics[scale=0.17]{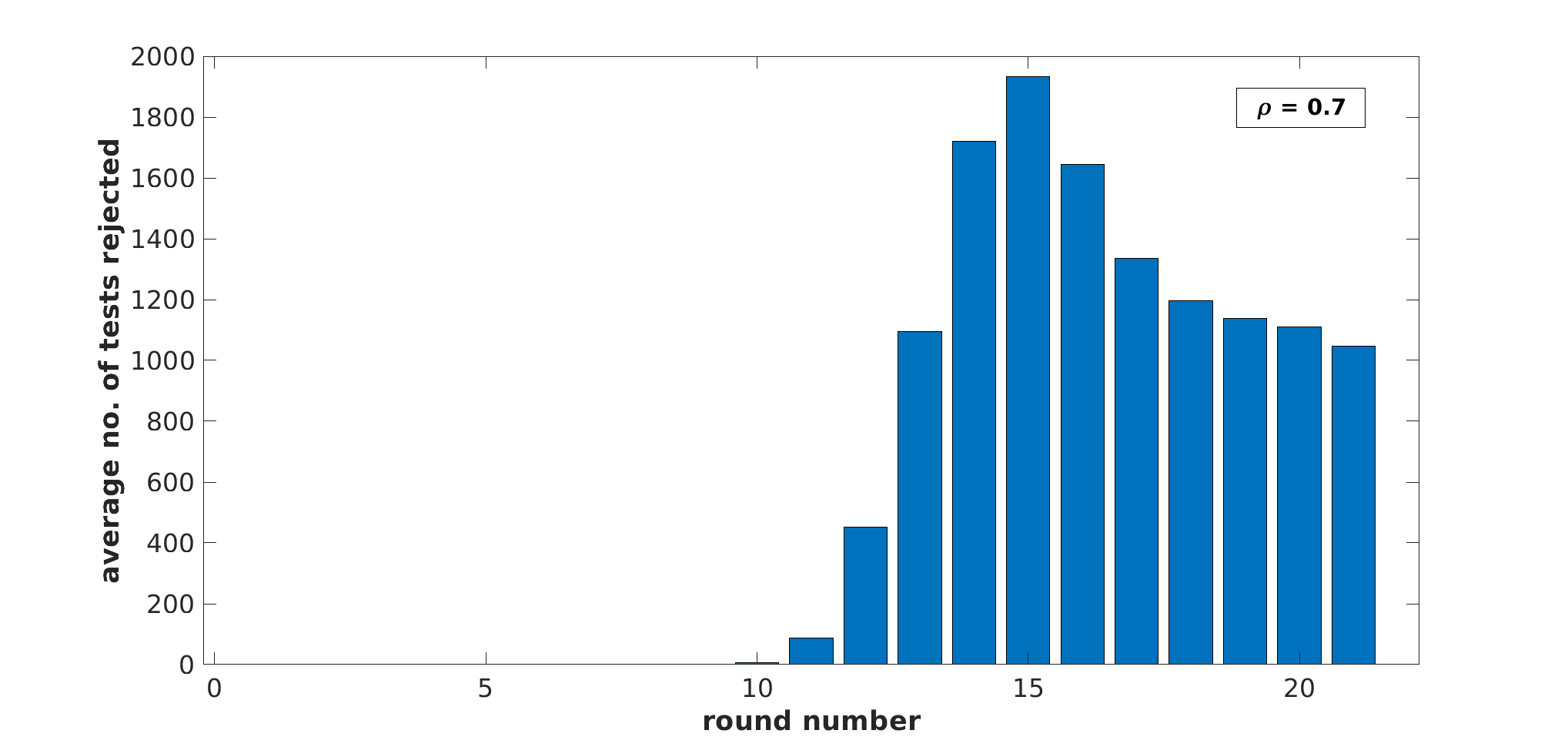}
    \includegraphics[scale=0.17]{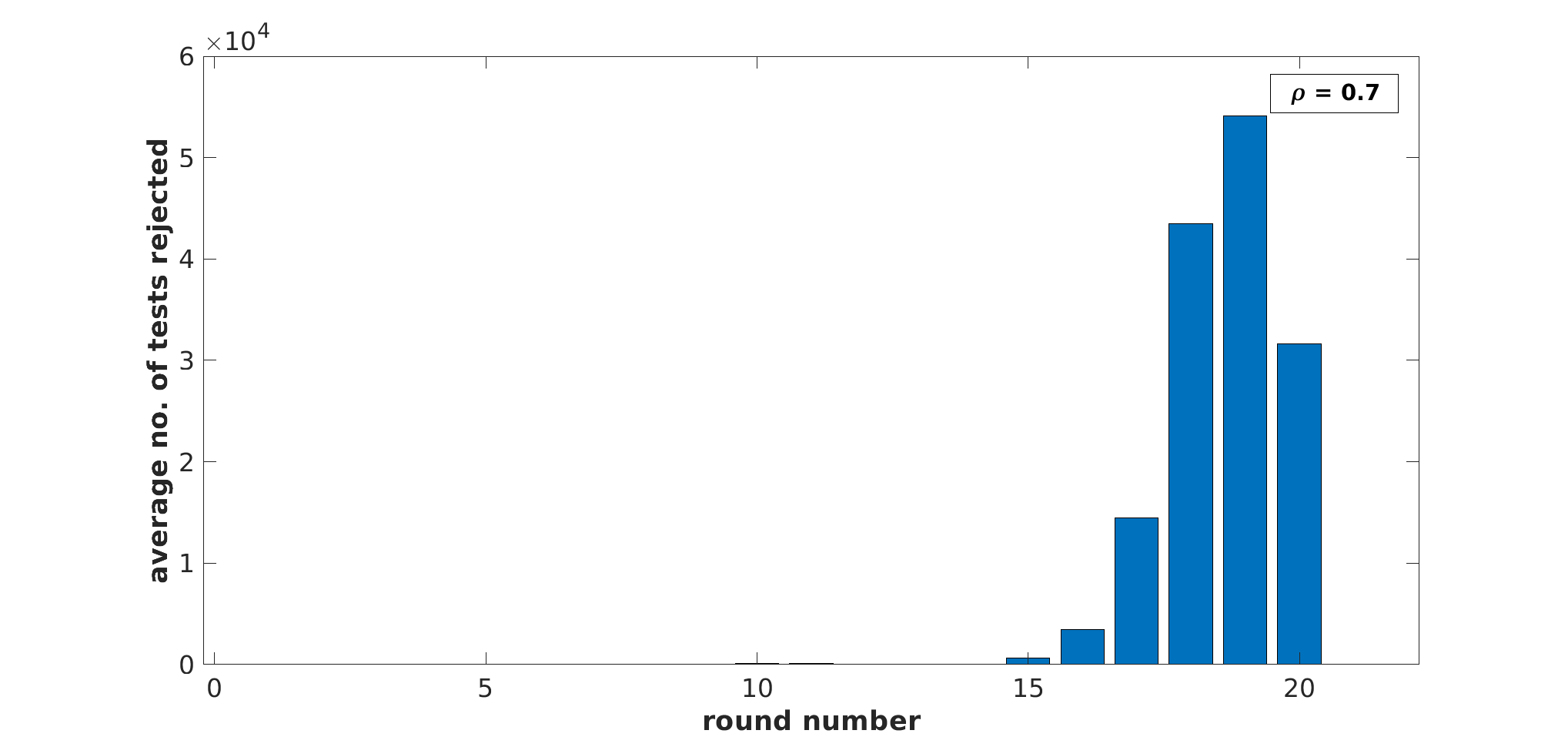}
    \includegraphics[scale=0.17]{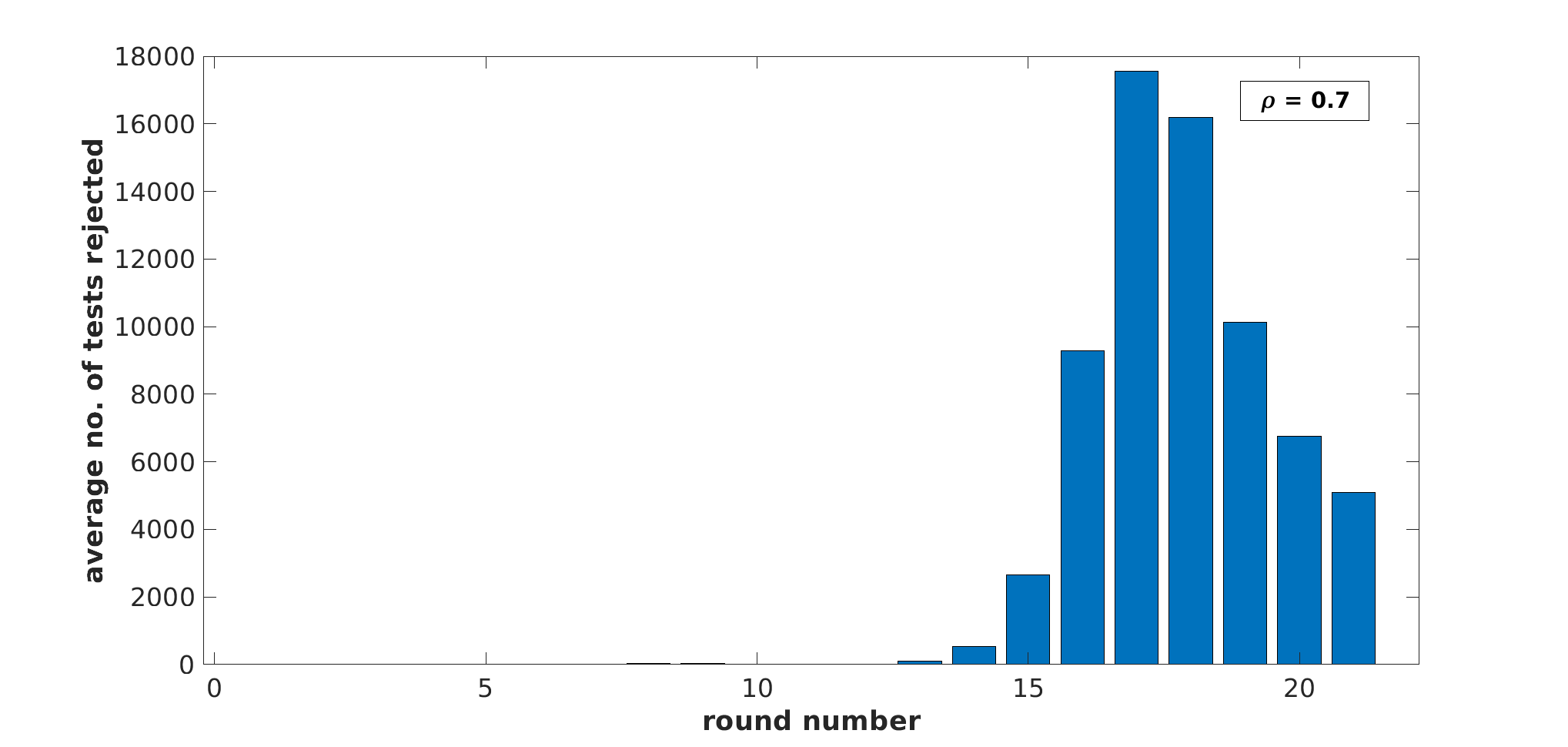}
    \caption{Histograms of the number of pools pruned in every round of binary splitting for the \textsc{Our-Sum} method, for \textsf{MIRFLICKR}, \textsf{ImageNet}, \textsf{IMDB-Wiki} and \textsf{InstaCities} (left to right, top to bottom), all for $\rho \geq 0.7$.}
    \label{fig:rejection_round_sum}
\end{figure*}

\begin{figure*}
    \centering
    \includegraphics[scale=0.35]{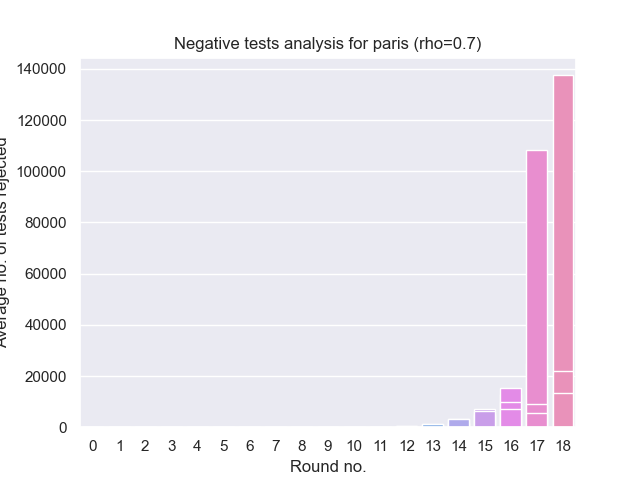}
    \includegraphics[scale=0.35]{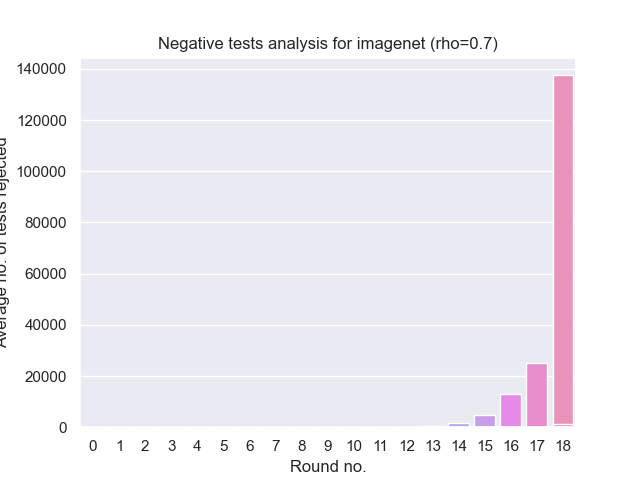}
    \includegraphics[scale=0.35]{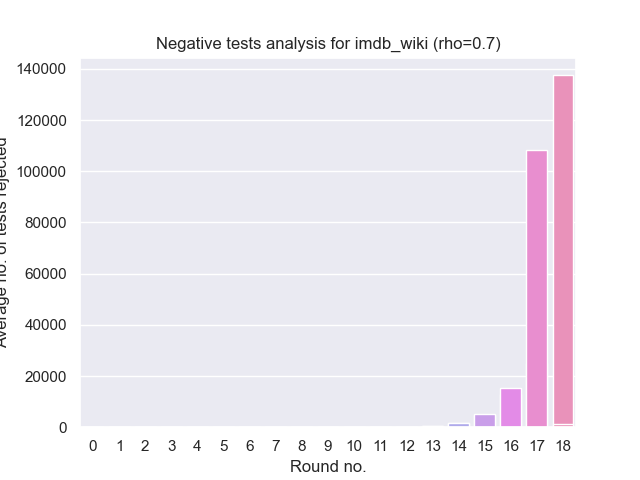}
    \includegraphics[scale=0.35]{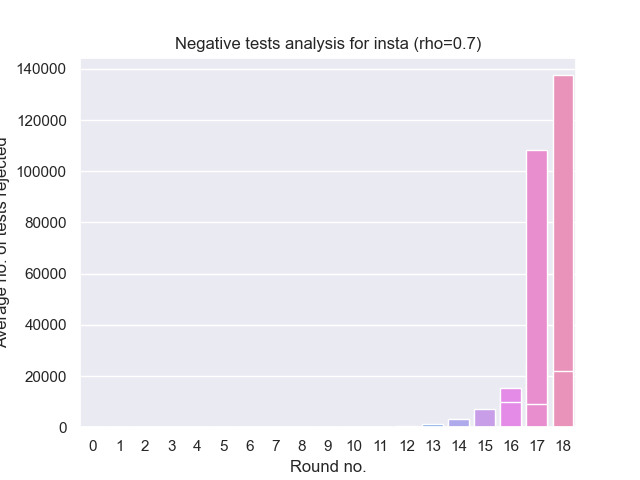}
    \caption{Histograms of the number of pools pruned in every round of binary splitting for the \textsc{Our-Max} technique for \textsf{MIRFLICKR}, \textsf{ImageNet}, \textsf{IMDB-Wiki} and \textsf{InstaCities} (left to right, top to bottom), all for $\rho \geq 0.7$.}
    \label{fig:rejection_round_max}
\end{figure*}
\begin{figure*}
    \centering
    \includegraphics[scale=0.25]{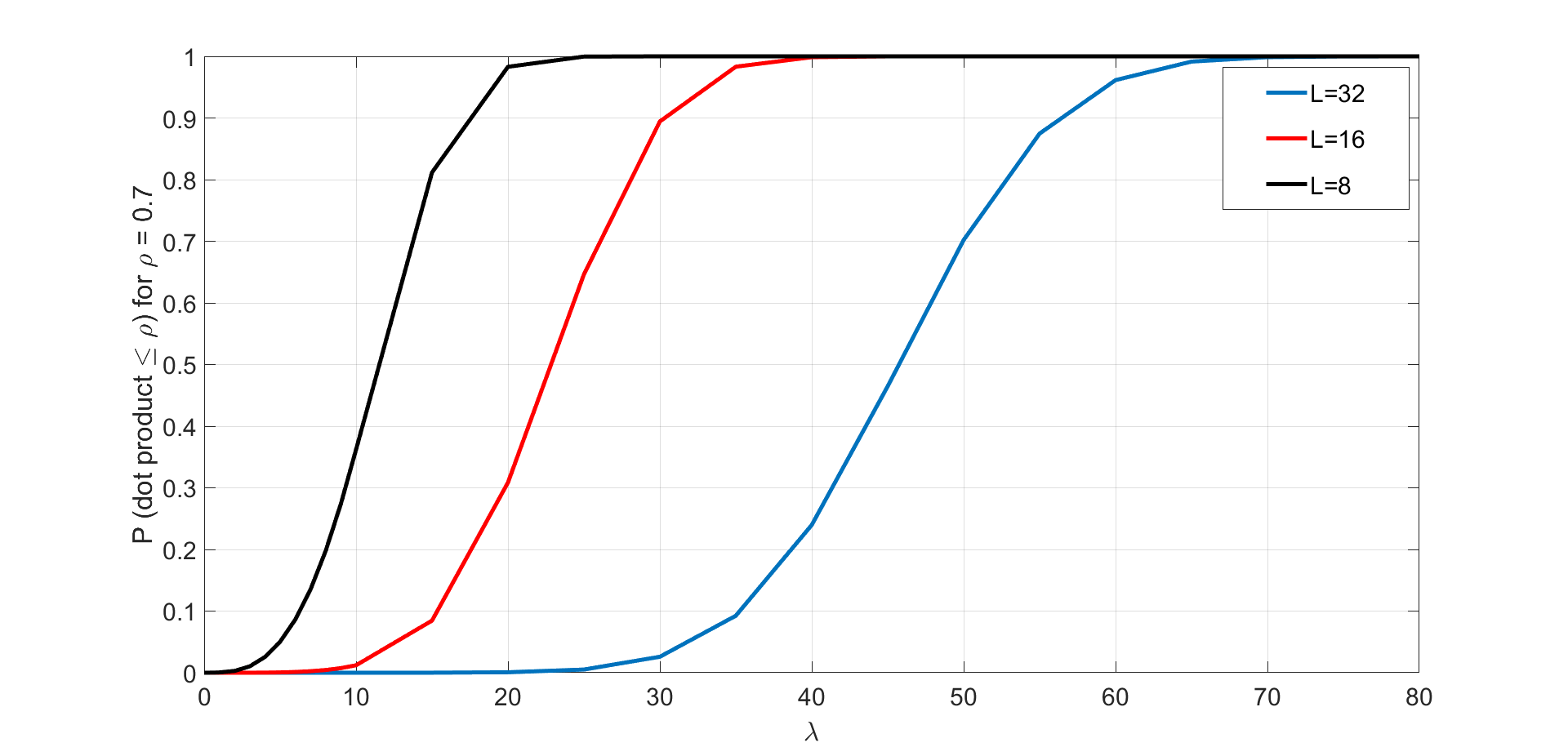}
    \caption{The probability that the dot product of a query vector $\boldsymbol{q}$ with a pool created from $L$ participating vectors, falls below $\rho = 0.7$, as a function of $\lambda$ for $L \in \{8,16,32\}$.}
    \label{fig:prob_dotproducts}
\end{figure*}

\section{Comparison to other Group Testing Methods for NN Search}
\label{sec:comparison_gt}
The oldest work on GT for NN search was presented in \cite{Shi2014}. In this method, the original data vectors are represented as a matrix $\boldsymbol{F} \in \mathbb{R}^{N \times d}$. The pools are represented as a matrix $\boldsymbol{Y} \in \mathbb{R}^{m \times d}$ obtained by pre-multiplying $\boldsymbol{F}$ by a randomly generated balanced binary matrix $\boldsymbol{A}$ of size $m \times N$ where $m < N$. 
Given a query vector $\boldsymbol{q}$, its similarity with the $j$th pool is computed, yielding a score $v_{yj} = \boldsymbol{q}^t \boldsymbol{Y^j}$, where $\boldsymbol{Y^j}$ is the $j$th group vector (and the $j$th row of $\boldsymbol{Y}$). This is repeated for every $j \in [m]$. Let $\mathcal{P}_i$ be the set of pools in which the $i$th vector, i.e. $\boldsymbol{f_i}$, belongs. Then, a likelihood score $L_i \triangleq \sum_{j \in \mathcal{P}_i} v_{yj}$ is computed for every vector. The vector from $\boldsymbol{F}$ with the largest likelihood score (denote this vector by $\boldsymbol{f_{k0}}$) is considered to be one of the nearest neighbors of $\boldsymbol{q}$. Next, the value $\boldsymbol{q}^t \boldsymbol{f_{k0}}$ is computed and the pool-level similarity scores for all pools containing  $\boldsymbol{f_{k0}}$ are updated to yield new scores of the form $v_{yj} = v_{yj} - \boldsymbol{q}^t \boldsymbol{f_{k0}}$. Thereafter, all likelihood scores are updated. This process is called back-propagation in \cite{Shi2014}. Again, the vector from $\boldsymbol{F}$ with the highest likelihood is identified and this procedure is repeated some $R$ times, and finally a sort operation is performed in order to rank them. In \cite{Shi2014}, it is argued that the time complexity of this procedure is $O( d (m + R))$. However, this is an approximate search method, and there is no guarantee that the $R$ neighbors thus derived will indeed be the nearest ones. As a result, a large number of false positives may be generated by this method. 

The work in \cite{Iscen2018} clusters the collection $\mathcal{D}$ into disjoint groups such that the members of a group are as orthogonal to each other as possible. As a result, the similarity measure $\boldsymbol{q}^t \boldsymbol{y_j}$ between $\boldsymbol{q}$ and a group vector $\boldsymbol{y_j}$ is almost completely dominated by the similarity between $\boldsymbol{q}$ and exactly one member of the group. A careful sparse coding and decoder correction step is also used, and the sparsity of the codes plays a key role in the speedup achieved by this method. However, this method is again an approximate method and heavily relies on the near-orthogonality of each group. 

The work in \cite{Engels2021} randomly distributes the collection $\mathcal{D}$ of $N$ vectors over some $B$ cells. This is independently repeated $R$ times, creating a grid of $B \times R$ cells. Within each cell, we have a collection of participating members $M_{r,b}$ and a corresponding group test $C_{r,b}$. The group test is essentially a binary classifier which determines whether $M_{r,b}$ contains at least one member which is similar to the query vector $\boldsymbol{q}$. This is an approximate query, with a true positive rate $p$ and false positive rate $q$. It is implemented via a distance-sensitive bloom filter \cite{Bloom1970,Kirsch2006}, which allows for very fast querying. The bloom filter is constructed with $m$ binary arrays, with some $L$ concatenated hash functions used in each array. For all positive tests $C_{r,b}$, a union set of all their members is computed. This is repeated $R$ times to create $R$ candidate sets, and the intersection of these $R$ sets is created. Each union can contain many false positives, but this intersection filters out non-members effectively. Precise bounds on $p,q$ are derived in \cite{Engels2021}, and the time complexity of a single query is proved to be $O(N^{1/2+\gamma} \log^3 N)$ where $\gamma \triangleq \log s_{|K|}/(\log s_{|K|+1} - \log s_{|K|})$ where $s_{|K|}$ stands for the similarity between $\boldsymbol{q}$ and the $K$th most similar vector in $\mathcal{D}$. The query time is provably sub-linear for queries for which $\gamma < 1/2$. This is obeyed in data which are distributed as per a Gaussian mixture model with components that have well spread out mean vectors and with smaller variances. But for many distributions, this condition could be violated, leading to arbitrarily high query times in the worst case. This method, too, produces a large number of false negatives and false positives. 

Compared to these three afore-mentioned techniques, our method is exact, with the same accuracy as exhaustive search. The methods \cite{Shi2014,Iscen2018,Engels2021} require hyper-parameters ($R$ for \cite{Shi2014}, sparse coding parameters in \cite{Iscen2018}, $B, R, m, L$ in \cite{Engels2021}) for which there is no clear data-driven selection procedure. Our method, however, requires no parameter tuning. In experiments, we have observed a speed up of more than ten-fold in querying time with our method as compared to exhaustive search on some datasets. Like \cite{Shi2014}, and unlike \cite{Engels2021, Iscen2018}, our method does have a large memory requirement, as we require all pools to be in memory. Some techniques such as \cite{Engels2021} require the nearest neighbors to be significantly more similar than all other members of $\mathcal{D}$ (let us call this condition $\mathscr{C}1$). Our method does not have such a requirement. However, our method will perform more efficiently for queries for which a large number of similarity values turn out to be small, and only a minority are above the threshold $\rho$ (let us call this condition $\mathscr{C}2$). In our experimental study on diverse datasets, we have observed that $\mathscr{C}2$ is  true always. On the other hand, $\mathscr{C}1$ does not hold true in a large number of cases, as also reported in \cite[Sec. 5.3]{Engels2021}.

\section{A Discussion on using Compressed Sensing Techniques for NN Search}
\label{sec:CS_limitations}
Sec. 3 of the main paper describes three recent papers \cite{Shi2014,Iscen2018,Engels2021} which apply the principles of group testing to near neighbor search, and compares them to our technique. In this section, we describe our attempts to apply the latest developments from the compressed sensing (CS) literature to this problem. Note that CS and GT are allied problems, and hence it makes sense to comment on the application of CS to near neighbor search. 

Consider that the original data vectors are represented as a matrix $\boldsymbol{F} \in \mathbb{R}^{N \times d}$. The pools are represented as a matrix $\boldsymbol{Y} \in \mathbb{R}^{m \times d}$ obtained by pre-multiplying $\boldsymbol{F}$ by a randomly generated balanced binary matrix $\boldsymbol{A}$ of size $m \times N$ where $m < N$. We considered the relation $\boldsymbol{v_y} = \boldsymbol{A v_x}$, where $\boldsymbol{v_y} \in \mathbb{R}^m$ contains the dot products of the query vector $\boldsymbol{q}$ with each of the pool vectors in $\boldsymbol{Y} \in \mathbb{R}^{m \times d}$, i.e. $\boldsymbol{v_y} = \boldsymbol{Yq}$. Likewise $\boldsymbol{v_x} \in \mathbb{R}^N$ contains the dot products of the query vector $\boldsymbol{q}$ with each of the vectors in the collection $\mathcal{D}$, i.e. $\boldsymbol{v_x} = \boldsymbol{Fq}$. The aim is to efficiently recover the largest elements of $\boldsymbol{v_x}$ from $\boldsymbol{A}, \boldsymbol{v_y}$. Since algorithms such as \textsc{Lasso} \cite{Hastie2015} are of an iterative nature, we considered efficient non-iterative algorithms from the recent literature for recovery of nearly sparse vectors \cite[Alg. 8.3]{Vidyasagar2019}. These are called expander recovery algorithms. Theorem 8.4 of \cite{Vidyasagar2019} guarantees recovery of the $k$ largest elements of $\boldsymbol{v_x}$ using this algorithm, but the bounds on the recovery error are too high to be useful, i.e. $\|\boldsymbol{v^{(k)}_x}-\boldsymbol{v_{x,est}}\|_{\infty} \leq \delta \triangleq \|\boldsymbol{v_x}-\boldsymbol{v^{(k)}_x}\|_1$. Here, the vector $\boldsymbol{v^{(k)}_x}$ is constructed as follows: the $k$ largest elements of $\boldsymbol{v_x}$ are copied into $\boldsymbol{v^{(k)}_x}$ at the corresponding indices, and all other elements of $\boldsymbol{v^{(k)}_x}$ are set to 0. Clearly, for most situations, $\delta$ will be too large to be useful. Indeed, in some of our numerical simulations, we observed $\delta$ to be several times larger than the sum of the $k$ largest elements of $\boldsymbol{v_x}$, due to which Alg. 8.3 from \cite{Vidyasagar2019} is rendered ineffective for our application. Moreover, for Theorem 8.4 of \cite{Vidyasagar2019} to hold true, the matrix $\boldsymbol{A}$ requires each item in $\mathcal{D}$ to participate in a very large number of pools, rendering the procedure inefficient for our application. Given these restrictive conditions, we did not continue with this approach. 

\section{Theoretical Analysis for \textsc{Our-Max}}
\label{sec:theory_our_max}
In Sec. 4 of the main paper, we have presented a theoretical analysis of the expected number of dot product computations required for \textsc{Our-Sum}, which is the binary splitting procedure using summations for pool creation. A similar analysis for estimating the number of dot products required for \textsc{Our-Max} using solely the distribution assumption on dot products (truncated normalized exponential or TNE) is challenging. 
Consider a pool vector $\boldsymbol{y} = \sum_{i=1}^K \boldsymbol{f_i}$ created by the summation of $K$ participating vectors $\{\boldsymbol{f_i}\}_{i=1}^K$. In \textsc{Our-Sum}, the dot product between the query vector $\boldsymbol{q}$ and the pool vector $\boldsymbol{y}$ is exactly equal to the sum of the dot products between $\boldsymbol{q}$ and individual members of the pool. Hence an analysis via the central limit theorem or the truncated Erlang distribution is possible, as explained in Sec. 5 of the main paper. For \textsc{Our-Max}, the pool vector $\boldsymbol{y}$ is constructed in the following manner: $\forall j \in \{1,2,...,d\}, y_j = \textrm{max}_{i \in \{1,2,...,K\}} f_{i,j}$ where $f_{i,j}$ stands for the $j$th element of vector $\boldsymbol{f_i}$. The dot product $\boldsymbol{q}^t \boldsymbol{y}$ is an \emph{upper bound} on $\textrm{max}_{_{i \in \{1,2,...,K\}}} \boldsymbol{q}^t \boldsymbol{f_i}$ (assuming that all values in $\boldsymbol{q}$ and in each $\boldsymbol{f_i}$ are non-negative). One can of course use the TNE assumption of the individual dot products $\boldsymbol{q}^t \boldsymbol{f_i}$ and use analytical expressions for the maximum of TNE random variables. However the fact that $\boldsymbol{q}^t \boldsymbol{y}$ is an \emph{upper bound} on $\textrm{max}_{_{i \in \{1,2,...,K\}}} \boldsymbol{q}^t \boldsymbol{f_i}$ complicates the analysis as compared to the case with \textsc{Our-Sum}. Hence, it is difficult to determine an expected number of dot-products per query for \textsc{Our-Max}, but it is possible to determine an upper bound on this quantity. This can be done as described below.

Let $X_i \triangleq \boldsymbol{q}^t \boldsymbol{f_i}$. Let $D_P$ denote the similarity of the query vector $\boldsymbol{q}$ with the pool vector $\boldsymbol{y}$, and let $P$ denote the set of vectors that participated in that pool. We can bound $D_P$ with a constant $c$ such that $D_P \leq c\cdot \text{max}_{i\in P}(X_i)$ (see later in this section regarding the value of $c$). Let $n := |P|$. Then using this bound, we have the following relation:
\begin{equation}
p_n(\rho) = P[D_P < \rho] \geq P[c \cdot \text{max}_{i\in P}(X_i) < \rho] \geq [F_X\left(\rho/c\right)]^{n}, 
\label{eq:d_bound2}
\end{equation}
where the last inequality is based on order statistics, and where $F_X(.)$ denotes the CDF of the TNE. From the main paper, recall that $Q_k$ stands for the expected number of pools in round $k$. Using the recursive relation for $Q_k$ in terms of $Q_{k-1}$, we have the following:
\begin{equation}
Q_k \leq 2(1 - [F_X\left(\rho/c\right)]^{N/2^{k-2}})Q_{k-1}. 
\label{eq:d_bound4}
\end{equation}
Using this, an upper bound on $E(\rho)$ can be obtained in the following manner:
\begin{equation}
E(\rho) \triangleq 1+\sum\limits_{i=2}^{\lceil \log_2 N +1 \rceil} Q_i.
\label{eq:E_rho2}
\end{equation}
Notice there is no $\frac{1}{2}$ factor (in contrast to $E(\rho)$ for \textsc{Our-Sum}) here. This is because even after splitting, the similarity needs to be calculated for both the divided pools unlike in the \textsc{Our-sum} algorithm. The constant $c$ can be theoretically as large as the dimension of the dataset (that is, $d = 1000$ for the datasets used here), giving a very loose upper bound. However, in order to obtain a sharper upper bound, we find the distribution of $c$ for each pool in a given dataset and use the 90 percentile value (denoted by $c_{90}$) to evaluate the bound. Details for this are provided next.

\begin{figure*}[h!]
    \centering
    \includegraphics[scale=0.35]{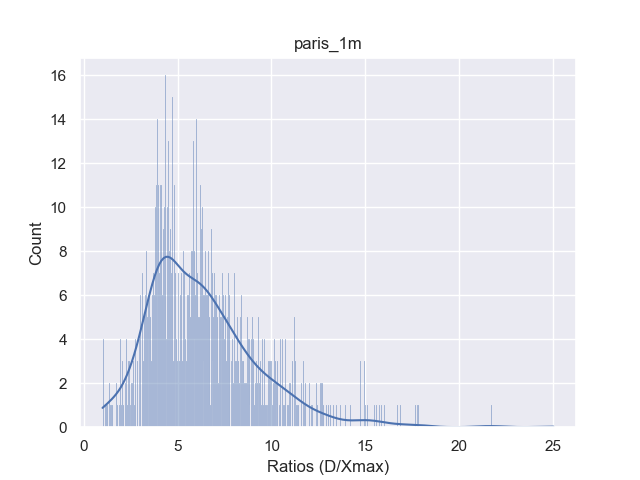}
    \includegraphics[scale=0.35]{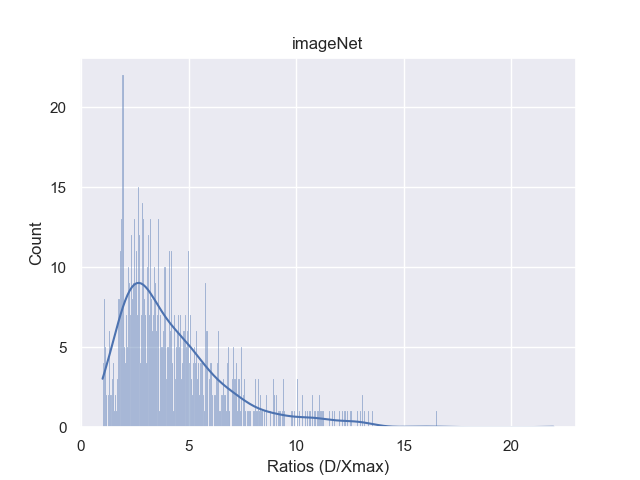}
    \includegraphics[scale=0.35]{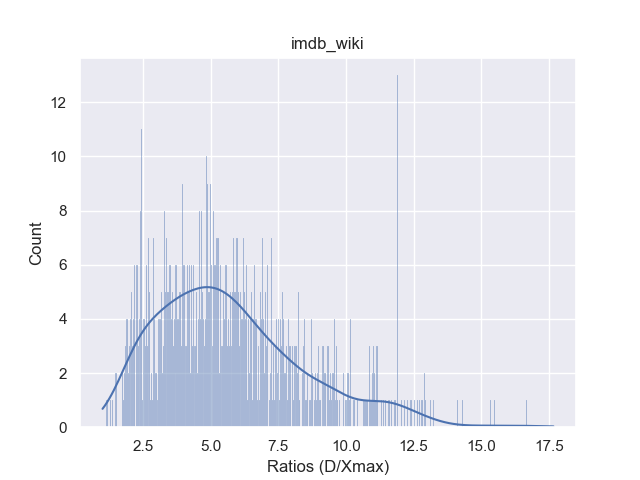}
    \includegraphics[scale=0.35]{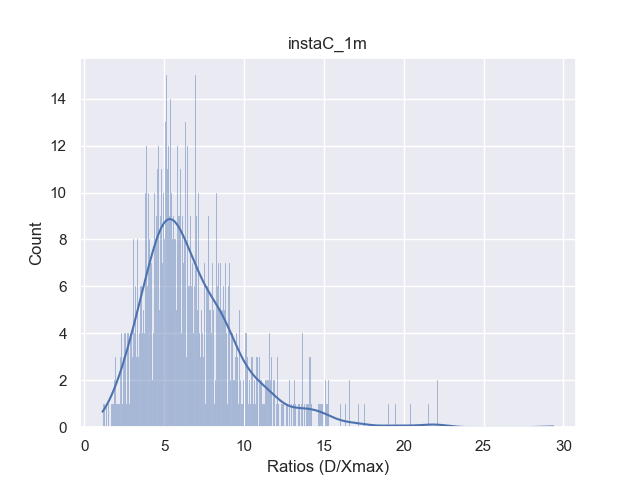}
    \caption{Histograms showing distribution of the ratio $c=\frac{D_{P}}{\text{max}_{i\in P}(X_i)}$ for pools in datasets \textsf{MIRFLICKR}, \textsf{ImageNet}, \textsf{IMDB-Wiki} and \textsf{InstaCities} (left to right, top to bottom). The histograms are computed over $\sim 1000$ queries, with 100 bins per histogram.}
    \label{fig:c_vals}
\end{figure*}

\noindent\textbf{Distribution of ratios $\left(c=\frac{D_P}{\textrm{max}_{i\in P}(X_i)}\right)$}:
Any upper bound on the expected number of dot products required by \textsc{Our-Max} is dependent on the constant $c$. Although, theoretically $c$ can be as large as the dimension of the database vectors, the distribution of $c$ for each dataset studied and 90 percentile value of $c$ (i.e., $c_{90}$) can be used to obtain the upper bounds on $E[\rho]$. Fig.~\ref{fig:c_vals} shows the distribution of $c$ for the datasets used in our experiments. Here below, we mention the $c_{90}$ values for various datasets used in evaluating the theoretical upper bound reported in Table 2 of the main paper.
\begin{itemize}
    \item \textsf{MIRFLICKR}: $c_{90} \approx 10$.
    \item \textsf{ImageNet}: $c_{90} \approx 7$.
    \item \textsf{IMDB-Wiki}: $c_{90} \approx 10$.
    \item \textsf{InstaCities}: $c_{90} \approx 11$.
\end{itemize}
Clearly, these values are significantly lower than $d$, which facilitates better upper bounds, albeit with some probability of error.

\section{Hyper-parameters for the algorithms}
\label{sec:hyperparameter}
In order to arrive at comparable values of precision and recall of other algorithms as compared to our proposed algorithms, we performed experiments with many hyper-parameters for each algorithm. Below are the hyper-parameters we used for various experiments with the competing methods. Note that in order to select the set of hyperparameters for each algorithm, we chose the one which gives high recall without significantly increasing time. So there can be other hyperparameter sets for an algorithm with higher recall but at the cost of disproportionate increase in query time. Both streaming and non-streaming setting uses the same set of hyperparameters for each algorithm, except \textsc{Flann-R} (details mentioned below). 
\begin{enumerate}
\item \textsc{Flinng} \cite{Engels2021}: For each dataset, we ran this code with different hyper-parameters $B \in \{2^{12}, 2^{13}, \cdots, 2^{18}, 2^{19}\}$, $m \in \{2^2,2^3,\cdots,2^{10}\}$, $R \in \{2, 3, 4\}, L = 14$ as recommended in the suppl. mat. of \cite{Engels2021} (see Sec. 3 of the main paper for their precise meaning) and chose the result that yielded the best recall.

\item \textsc{Falconn} \cite{Andoni2015}:
Number of tables = $\{100, 150, 200, 250, 300\}$, Number of probes = $\{40, 50, 70, 100\}$. Parameters used for reporting : (250, 70) respectively 

\item \textsc{Ivf} from the \textsc{Faiss} package \cite{Faiss,Babenko2014}: Number of lists = $\{32, 64, 128\}$, Number of probes = $\{2, 16\}$. Parameters used for reporting : (32, 16) respectively

\item \textsc{Hsnw} from the \textsc{Faiss} package \cite{Faiss}: $M = 32$, Efconstruction = $\{32, 64\}$, Efsearch = $\{2\}\times\text{Number of neighbors to retrieve}$, i.e. $K$. Parameters used for reporting : (32, 2) respectively

\item \textsc{Scann} \cite{scann,Guo2020}:
Number of leaves = $\{1,2\} \times \sqrt{\text{Number of elements in the dataset}}$, Number of leaves to search = $\{1/2, 1/4, 1/8\}\times\text{Number of leaves}$, Reorder number of neighbors = $\{4, 8, 16\} \times K$. Parameters used for reporting : (2, 1/4, 16) respectively

\item \textsc{Flann} \cite{Muja2014,FlannPy}: We used \texttt{AutotunedIndexParams} with high target precision (0.85) and other parameters set to their default values. For IMDB-Wiki dataset, we used \texttt{KDTreeIndexParams(32)}, for building the index with 32 parallel kd-trees and \texttt{SearchParams(128)} during search phase. This was done because \texttt{AutotunedIndexParams} took long time to build index (more than 20 hours) on IMDB-Wiki dataset. For streaming setting, due to excess time taken by \texttt{AutotunedIndexParams} (leading to high \textit{Index Build Time}), we used \texttt{KDTreeIndexParams(16)}, for building the index with 16 parallel kd-trees and \texttt{SearchParams(256)}.

\item \textsc{Falconn++} \cite{falcon_plus_plus,pham2022}: 
We used the parameters suggested in examples of \textsc{Falconn++}, with number of random vectors = 256, number of tables = 350, $\alpha = 0.01$, Index Probes (iProbes) = 4, 
Query Probes (qProbes) = $\textrm{min}(40k, 2\cdot\textrm{max}(k \text{  among all queries})$. But we have different values of $k$ for each query which can be as large as $N/10$. Hence, in order to prevent qProbes from exceeding the size of dataset, we cap it to $2\cdot\textrm{max}(k \text{ among all queries})$). 
\end{enumerate}
\renewcommand{\arraystretch}{1.5}
\setlength{\tabcolsep}{8pt}
\begin{table}[h!]
\fontsize{7pt}{7pt}\selectfont
\centering
\begin{subtable}[h]{0.9\textwidth}
    \begin{tabular}{|c|c|c|c|c|c|c|c|c|}
    \hline
    \rowcolor{Gray} 
    \multicolumn{9}{|c|}{ImageNet}
    \\\hline
    \rowcolor{Gray} 
    & \multicolumn{4}{|c|}{Precision $\uparrow$} &
    \multicolumn{4}{|c|}{Recall $\uparrow$}\\\hline
    \rowcolor{Gray} 
    Feature$\downarrow$, $\rho \rightarrow$ & $0.6$ & $0.7$ & $0.8$ & $0.9$ & $0.6$ & $0.7$ & $0.8$ & $0.9$ \\
    \hline
    Softmax & 0.7 & 0.73 & 0.76 & 0.81 & 0.74 & 0.70 & 0.66 & 0.60\\\hline
    Penultimate & 0.66 & 0.85 & 0.96 & 0.99 & 0.2 & 0.07 & 0.015 & 0.001\\
    \hline
    \end{tabular}
\end{subtable}
\vspace{0.4cm}

\begin{subtable}[h]{0.9\textwidth}
    \begin{tabular}{|c|c|c|c|c|c|c|c|c|}
    \hline
    \rowcolor{Gray} 
    \multicolumn{9}{|c|}{ImageDBCorel}
    \\\hline
    \rowcolor{Gray} 
    & \multicolumn{4}{|c|}{Precision $\uparrow$} &
    \multicolumn{4}{|c|}{Recall $\uparrow$}\\\hline
    \rowcolor{Gray} 
    Feature$\downarrow$, $\rho \rightarrow$ & $0.6$ & $0.7$ & $0.8$ & $0.9$ & $0.6$ & $0.7$ & $0.8$ & $0.9$ \\
    \hline
    Softmax & 0.95 & 0.96 & 0.97 & 0.98 & 0.32 & 0.27 & 0.23 & 0.18\\\hline
    Penultimate & 0.99 & 1 & 1 & 1 & 0.29 & 0.11 & 0.028 & 0.01\\
    \hline
    \end{tabular}
\end{subtable}
\vspace{0.4cm}

\begin{subtable}[h]{0.9\textwidth}
    \begin{tabular}{|c|c|c|c|c|c|c|c|c|}
    \hline
    \rowcolor{Gray} 
    \multicolumn{9}{|c|}{ImageDBCaltech}
    \\\hline
    \rowcolor{Gray} 
    & \multicolumn{4}{|c|}{Precision $\uparrow$} &
    \multicolumn{4}{|c|}{Recall $\uparrow$}\\\hline
    \rowcolor{Gray} 
    Feature$\downarrow$, $\rho \rightarrow$ & $0.6$ & $0.7$ & $0.8$ & $0.9$ & $0.6$ & $0.7$ & $0.8$ & $0.9$ \\
    \hline
    Softmax & 0.58 & 0.66 & 0.73 & 0.82 & 0.32 & 0.29 & 0.25 & 0.21\\\hline
    Penultimate & 0.89 & 0.97 & 0.99 & 0.99 & 0.19 & 0.08 & 0.03 & 0.02\\
    \hline
    \end{tabular}
\end{subtable}
     \vspace{0.4cm}
    \caption{Comparison of the retrieval accuracy (precision and recall) using cosine distance with (i) softmax features of VGGNet (1000 dimensional) and (ii) penultimate VGGNet features (4096 dimensional) for different values of cosine distance threshold $\rho \in \{0.6,0.7,0.8,0.9\}$. Note the higher recall of softmax features.}
   \label{tab:softmax_VGGNet}
\end{table}

\section{Image Augmentations}
\label{sec:image_aug}
\begin{figure*}[h!]
    \centering
    \includegraphics[scale=0.39]{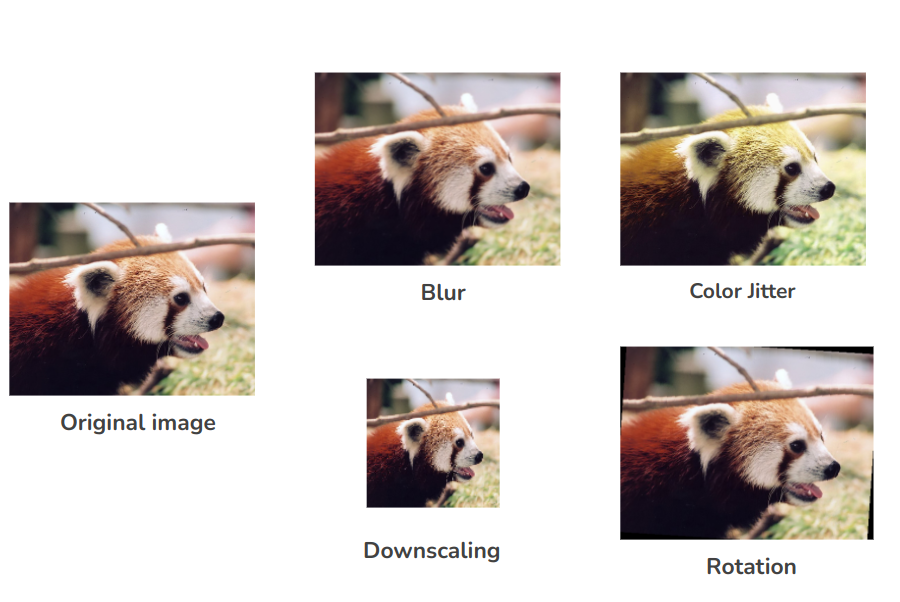}
    \caption{Example of augmentations performed on an image from the ImageNet dataset}
    \label{fig:augmentation}
\end{figure*}

In order to create queries suitable for the streaming environment for plagiarism detection, images from the database underwent specially tailored augmentations. The augmentations included artifacts that are somewhat subtle and which do not alter the image content very signficantly, but are designed to trick the detection system: 
Gaussian blur (with kernel size 3$\times$3), color jitter (in hue, saturation and value), horizontal flip, downscaling of images and image rotation (by 3 degrees). Figure \ref{fig:augmentation} shows an example of these augmentations.

\section{Comparison of VGG16 features}
\label{sec:comp_vgg16}
In this section, we provide a comparison between features extracted from penultimate layer of VGG16  (4096 dimensional) and the features extracted after the last layer (i.e. after applying the softmax functions leading to a 1000 dimensional feature vector as there are 1000 classes involved). For comparison, we use a setting akin to \cite{CBIR}, in which database images which are similar to a given query image are retrieved. However, instead of retrieving a fixed number of images ($k$), we have performed range-based retrieval with different similarity thresholds ($\rho=\{0.6, 0.7,0.8,0.9\}$) on ImageNet \cite{ImageNetKaggle}, ImageDBCaltech (Caltech101) \cite{CALTECH101} and ImageDBCorel \cite{COREL}. Table~\ref{tab:softmax_VGGNet} shows the retrieval recall and retrieval precision values for the aforementioned datasets. Note that we define 
\textbf{retrieval precision} = \# (images retrieved belonging to the same class as the query image) / \# (images retrieved), and 
\textbf{retrieval recall} = \# (images retrieved belonging to the same class as the query image) / \# (images in the database having the same class as the query image). 
Note that the retrieval precision and retrieval recall as defined here pertain to identification of the \emph{class labels}. These are different from the precision and recall reported in the main paper, which are based on near neighbors retrieved by a possibly approximate near neighbor search algorithm, in comparison to an exhaustive nearest neighbor search. Note that though the main algorithm proposed by our paper is fully accurate, the other algorithms we have compared to (i.e., \cite{Engels2021,Falconn_imp,Muja2014}) perform only approximate near neighbor search. 
Observe from Table~\ref{tab:softmax_VGGNet} that the recall rates for penultimate features noticeably lag behind those of softmax features, indicating that softmax features are better suited for plagiarism detection and other applications requiring high recall.

\section{Other Experiments}
\label{sec:other_exp}
Here we present results on two experiments with a static database: (\textit{i}) One on retrieval with low similarity values, i.e. $\rho=0.3$, and (\textit{ii}) One on the complete ImageNet Database containing 12 million data-points. These results are reported in Table~\ref{tab:results_additional}.

\setlength{\tabcolsep}{3pt}
\renewcommand{\arraystretch}{1.2}

\begin{table}[h]
\fontsize{9pt}{9pt}\selectfont
\begin{center}
\begin{tabular}{|c|c|c|c|c||c| }
 \hline
 \rowcolor{Gray}
 & \multicolumn{4}{|c|}{$\rho = 0.3$} & 10M scale, $\rho = 0.8$ \\ \hline
  \rowcolor{Gray}
 Alg./DB. & \textsf{ImgN.} & \textsf{IMDBW.} & \textsf{InstaC.} & \textsf{MIRFL.} & \textsf{ImgN. 12M}\\ 
 \hline
 \textsc{Ivf} (\textsc{Faiss}) & 157,0.99,1 & \textbf{51},0.91,1 & 97,0.97,1 &  98,0.98,1 & 1342,1,1 \\\hline
\textsc{Falconn} & 128,0.93,0.99 & 184,0.95,1 & 163,0.92,1 & 141, 0.89,1 & 
1521,0.96,1 \\\hline
\textsc{Hnsw} & 15,0.96,0.96 & 3126,0.97,0.97 & 432,0.97,0.97 & 223,0.98,0.98 & 561,0.98,0.98 \\\hline
\textsc{Our-Sum} & \textbf{7,1,1} & 133,\textbf{1,1} & \textbf{90,1,1} & \textbf{67,1,1} & \textbf{269,1,1}\\\hline
 \rowcolor{Gray}
\end{tabular}
\end{center}
\caption{Comparison of mean query times (ms), mean recall, mean precision (respectively) for different methods over 10K queries: (i) For $\rho = 0.3$ for different databases; and (ii) for the full ImageNet database with \textbf{12 million images} for $\rho = 0.8$. Hyperparameters for all algorithms are the same as mentioned earlier in in this supplemental material document. \textsc{Our-Sum} gives better precision, recall than all methods, and generally better query times than others.}
\label{tab:results_additional}
\end{table}

\end{document}